\documentclass[letterpaper,journal]{IEEEtran}
\usepackage{amsmath,amsfonts}
\usepackage{algorithmic}
\usepackage{algorithm}
\usepackage{array}
\usepackage[caption=false,font=normalsize,labelfont=sf,textfont=sf]{subfig}
\usepackage{textcomp}
\usepackage{stfloats}
\usepackage{url}
\usepackage{verbatim}
\usepackage{graphicx}
\usepackage{cite}
\usepackage{xcolor}
\usepackage{booktabs}
\usepackage{multirow}
\usepackage{makecell}
\usepackage{textcomp}
\usepackage[table]{xcolor} 
\newtheorem{prop}{Proposition}
\newcommand{\chold}[1]{\textcolor{black}{#1}}
\newcommand{\ch}[1]{\textcolor{black}{#1}}

\definecolor{forest}{RGB}{46, 139, 87}  
\definecolor{ruby}{RGB}{205, 92, 92}   
\definecolor{amethyst}{RGB}{153, 102, 204}
\colorlet{greenlight}{forest!15}
\colorlet{greenmed}{forest!25}
\colorlet{greendark}{forest!35}
\colorlet{redlight}{ruby!15}
\colorlet{redmed}{ruby!25}
\colorlet{reddark}{ruby!35}
\colorlet{purplelight}{amethyst!15}
\colorlet{purplemed}{amethyst!25}
\colorlet{purpledark}{amethyst!35}

\begin{document}

\title{Probability-Aware Parking Selection}

\author{Cameron Hickert,~\IEEEmembership{Member,~IEEE,} Sirui Li,~\IEEEmembership{Member,~IEEE,} Zhengbing He,~\IEEEmembership{Senior Member,~IEEE,} Cathy Wu,~\IEEEmembership{Member,~IEEE}
\thanks{This work was supported in part by Cintra, the MIT Energy Initiative (MITEI), and the National Science Foundation (NSF) under grant numbers 2239566 and 2434399. {\it Corresponding authors: Cameron Hickert,  Zhengbing He}}
\thanks{Cameron Hickert and Sirui Li are with the Laboratory for Information \& Decision Systems and the Institute for Data, Systems, and Society, Massachusetts Institute of Technology, Cambridge, MA 02139, USA {(Email: \tt\small chickert@mit.edu; siruil@mit.edu})}%
\thanks{Zhengbing He is with the Laboratory for Information \& Decision Systems, Massachusetts Institute of Technology, Cambridge, MA 02139, USA {(Email: \tt\small he.zb@hotmail.com})}%
\thanks{Cathy Wu is with the Laboratory for Information \& Decision Systems, the Institute for Data, Systems, and Society, and the Dept. of Civil and Environmental Engineering, Massachusetts Institute of Technology, Cambridge, MA 02139, USA {(Email: \tt\small cathywu@mit.edu})}}

\markboth{}%
{Shell \MakeLowercase{\textit{et al.}}: A Sample Article Using IEEEtran.cls for IEEE Journals}


\maketitle

\begin{abstract} 

\ch{Current navigation systems conflate time-to-drive with the true time-to-arrive by ignoring parking search duration and the final walking leg. Such underestimation can significantly affect user experience, mode choice, congestion, and emissions.} 
To address this issue, this paper introduces the probability-aware parking selection problem, which aims to direct drivers to the best parking location rather than straight to their destination.
An adaptable dynamic programming framework is proposed that leverages probabilistic, lot-level availability \ch{to minimize the expected time-to-arrive.}
Closed-form analysis determines when it is optimal to target a specific parking lot or explore alternatives, as well as the expected time cost.
Sensitivity analysis and three illustrative cases are examined, demonstrating the model’s ability to account for the dynamic nature of parking availability.
Given the high cost of permanent sensing infrastructure, we assess the error rates of using stochastic observations to estimate availability. Experiments with real-world data from the US city of Seattle indicate this approach's viability, with mean absolute error decreasing from 7\% to below 2\% as observation frequency increases. In data-based simulations, probability-aware strategies demonstrate time savings up to 66\% relative to probability-unaware baselines, yet still take up to 123\% longer than \ch{time-to-drive} estimates.

\end{abstract}


\section{Introduction}
\begin{figure}
	\center
	\includegraphics[width=0.48\textwidth, clip=true, trim=0 0 0 10]{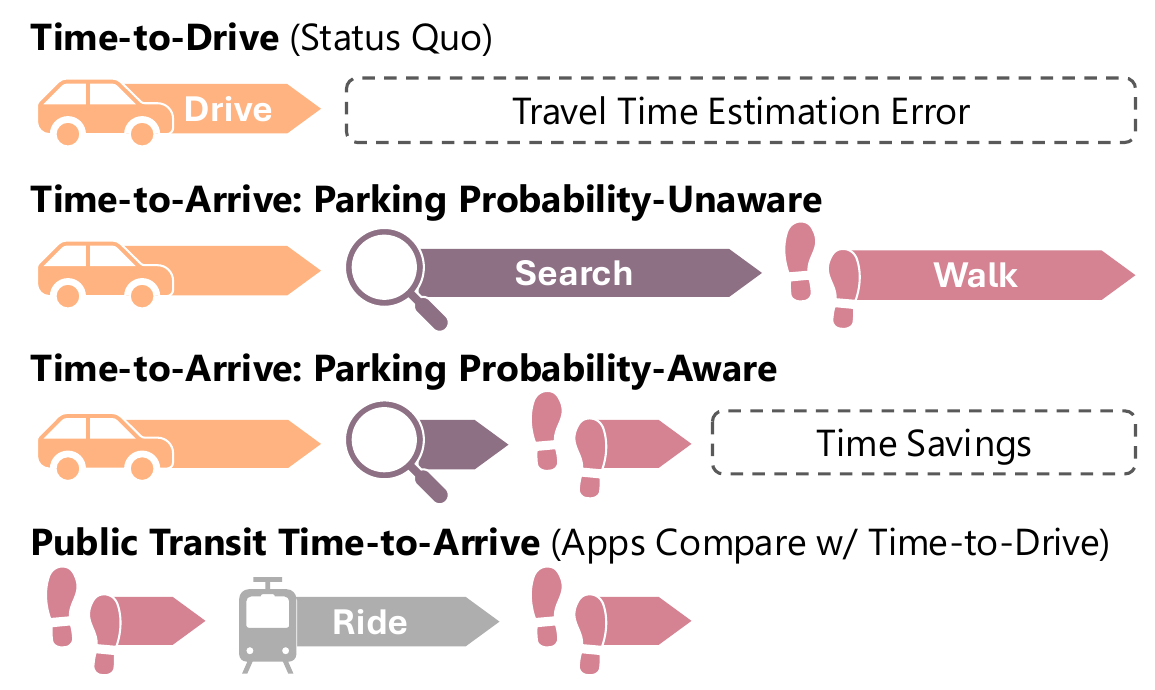}
	\caption{\ch{Accounting for the search for parking and the post-parking walking leg can increase the accuracy of personal vehicle travel time estimates, save time, and improve cross-mode comparability.}}
\label{fig:overview}
\end{figure}

\IEEEPARstart{C}{urrent} navigation apps typically send drivers to their destination, occasionally providing a basic assessment (e.g., ‘easy’, ‘difficult’) of parking in the area~\cite{arora2019hard}.
Given that parking spaces may not exist at the destination or may be unavailable, drivers often need to find parking elsewhere or cruise for parking at the current location until a spot becomes available. 
\chold{For example, consider the parking garage at San Francisco General Hospital, located in an area with limited on-street parking: when it is full, driving to and walking from the next-closest garage adds at least 30 minutes for patients and visitors, according to Google Maps routing estimates.}
Similar scenarios, familiar to many urban drivers and riders, highlight the frustrating difference between today's \ch{time-to-drive} estimates that ignore parking difficulties and more useful \ch{time-to-arrive} estimates that account for the parking search and post-parking walk time. \ch{See Figure~\ref{fig:overview}.}

Beyond inconvenience, underestimating the true time to arrive via personal vehicles may cause two other problems. First, it may prevent mode shift that would occur if individuals had more accurate estimates~\cite{arora2019hard}. 
This is particularly true given that popular navigation apps (e.g., Google Maps, Apple Maps) include estimates for walking time from public transit to a final destination in their total public transit travel time estimates, but do not include walking time from available parking to a final destination in their total drive time estimates \cite{Yafeng2024}. Second, cruising for parking is known to contribute to congestion and emissions in urban environments~\cite{shoup2006cruising}.

To address this issue, this paper introduces the probability-aware parking selection problem and proposes an adaptable dynamic programming framework to direct drivers to the best parking location based on location \textit{and} parking lot-level probabilistic availability information, rather than directly to their destination. 
The dynamic programming framework offers a structured and adaptable theoretical foundation for closed-form analysis of optimal parking strategies under uncertainty, sensitivity, changing probabilities, as well as a mechanism for understanding trade-offs involved. 
\chold{Results leveraging real-world parking data from Seattle indicate stochastic observations of true lot-level parking probabilities could be a viable estimation approach. 
In the analyzed cases, observed probabilities' mean absolute error begins below 7\% and falls below 2\% as observations become more frequent.}
\chold{Simulations with real-world data across different settings demonstrate the practical potential of probability-aware parking strategies. In all cases studied, these strategies outperform traditional methods, achieving time savings of up to 66\% in the setting with the highest congestion.}
The work makes \ch{four} contributions: 
\begin{itemize}
    \item \ch{It establishes time-to-arrive as a unified metric to enable fair comparison across transport modes, correcting the bias inherent in omitting search and walk times.}
    \item An introduction of the probability-aware parking selection problem and a corresponding adaptable dynamic programming solution.
    \item A closed-form analysis delineating when it is optimal to wait at a specific parking lot as opposed to when it may be better to visit other lots, as well as sensitivity analysis for dynamic probabilities.
    \item \ch{A demonstration of the time savings from} probability-aware parking strategies with stochastic observations. \ch{This includes} a comparative case study using real-world parking data from Seattle \ch{that illustrates the utility of time-to-arrive relative to time-to-drive}.
\end{itemize}

This work is organized as follows: Section~\ref{sec:related_work} reviews the related literature. Section~\ref{sec:methodology} introduces the dynamic programming problem and a corresponding solution, along with analysis for dynamic probability settings. Section~\ref{sec:experiments_stoch} details experiments that explore error from stochastic observations. \chold{Section~\ref{sec:seattle_sim} presents the Seattle case study}. In Section~\ref{sec:future_work} the paper sets forth implications for future work.

\section{Related Work}
\label{sec:related_work}
Probability-aware parking selection has key differences from the well-studied optimal stopping problem for parking~\cite{sakaguchi1982optimal, tamaki1982optimal, tamaki1988optimal}, as well as other theoretical analyses of cruising~\cite{arnott2017cruising}. Primarily, the driver can decide where to go to seek parking, rather than passively waiting for availability to appear along a pre-defined route with strictly decreasing distance (usually) to the destination. 
Our analysis also considers probabilities at the level of entire parking lots rather than individual spaces, since the former is easier to estimate~\cite{xiao2023parking}. Still, our framework can accommodate spot-level analysis by treating each as its own individual lot. Even in optimal stopping variants that allow backtracking, vehicles cannot alter the order of their visits to various spaces, the driving geometry is 1-dimensional, and parking availability odds rarely vary~\cite{tamaki1988optimal, krapivsky2019simple}. 

Two well-executed closely related works are those described in~\cite{djuric2016parkassistant} and~\cite{hedderich2018optimization}. \chold{Given information about a driver's parking preferences, the road network, traffic rules, and parking regulations, the algorithm in~\cite{djuric2016parkassistant} greedily calculates expected driver utility along possible routes to suggest a trajectory. \cite{hedderich2018optimization} describes a routing approach based on the A* algorithm with a hand-designed cost function and focuses on the optimization of that function's weight term, which is specific to a given road network and set of static parking probabilities.} However, these works consider only street parking, do not use closed-form analysis to express or compare possibilities, and focus on cruising within a predefined tolerance boundary of the final destination instead of \textit{a priori} selection of parking destinations separate from the final destination. \chold{Furthermore, neither formulates a dynamic programming framework for the problem, addresses how parking information may be obtained, nor considers dynamic parking probabilities.} 

Various mobile applications exist (e.g., SpotHero, SpotAngels, ParkCBR, Parking.com, Park Smarter) or have been proposed (e.g., ASPIRE~\cite{rizvi2018aspire}) for parking, but generally these are simple reservation and/or payment systems only for spots over which the app has control~\cite{teodorovic2006intelligent}. Other systems like iParker address pricing, but not probability-based selection~\cite{kotb2016iparker}.

The survey in~\cite{xiao2023parking} as well as the works in~\cite{bock2019smart} and~\cite{shi2018parkcrowd} investigate parking availability predictions. The present paper builds upon this research to better understand the utility of such enhanced parking estimates for users. \cite{wu2014agile} uses the possibility of time- and availability-aware parking selection to motivate the work, but focuses on empirical probability prediction at specific lots with in-built sensors in a vehicle's immediate vicinity. Still, our notion of parking probability aligns with theirs, i.e., the likelihood that at least one parking space is available when a vehicle arrives.

\begin{figure*}
	\center
	\includegraphics[width=0.85\textwidth, clip=true, trim=0 0 0 10]{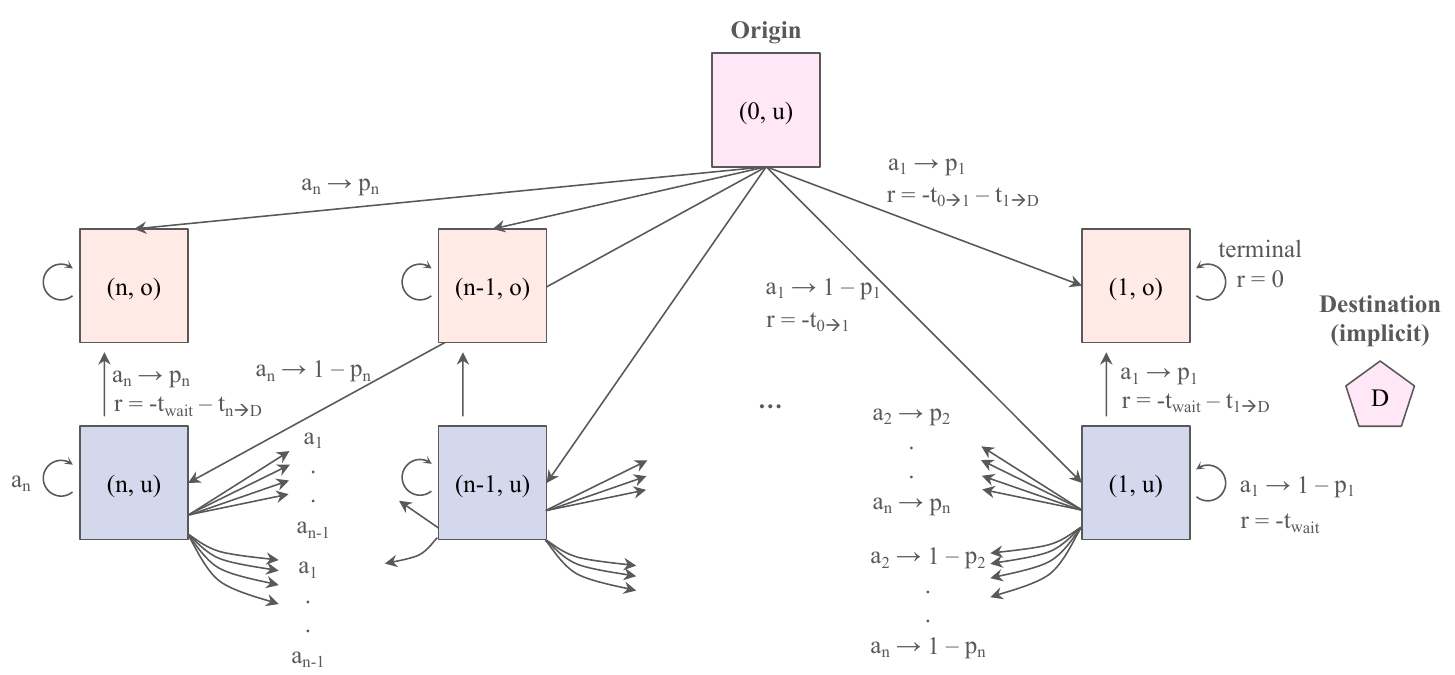}
	\caption{Visualization of the specified MDP. Orange states represent successful parking at the given lot $(\cdot, o)$, and are terminal. Purple states represent those where the driver visits a lot, cannot immediately park, and thus must decide whether to visit another lot or wait at the current lot $(\cdot, u)$. Actions ($a$) are shown with their associated probabilities ($p$), except where omitted for clarity of presentation. To avoid over-complication, only select rewards are shown; the full reward scheme is described in the main text.}
\label{fig:mdp_diagram}
\end{figure*}

\section{Methodology}
\label{sec:methodology}
\subsection{Framework}
\label{subsec:framework}

The initial challenge is to construct a framework capturing the spatial structure, time costs, information, and uncertainty involved in parking while enabling closed-form analysis. We model this situation using an infinite-horizon Markov decision process (MDP) $\mathcal{M} = (\mathcal{S}, \mathcal{A}, \mathcal{P}, \mathcal{R}, s_0)$; see Figure~\ref{fig:mdp_diagram}.
Each state $s \in \mathcal{S}$ is comprised of a tuple $(i, \{u, o\})$ where $i \in \{0, ..., N\}$ indexes the origin ($0$) and any of $N$ on- or off-street parking lots, and the set $\{u, o\}$ indicates the vehicle's parking status, where $u$ is \textit{u}nparked and $o$ is parked in an \textit{o}pen spot. 
The initial state $s_0 = (0, u)$ thus indicates an unparked vehicle at the origin. Parked states are terminal.

At each timestep, let action $a_i \in \mathcal{A} = \{ 1, \dots, N \}$ represent an attempt to park at lot $i$. 
Note that since there is no $a_0$, a vehicle cannot attempt to stay at or return to the origin. 
This attempt succeeds with a lot-specific probability $p_i \in \mathcal{P}$ and fails with probability $1 - p_i$. We assume $p_i \in (0, 1]$.
Let $r_{i, (j, \{u, o\})} \in \mathcal{R}$ represent the instantaneous reward incurred in the process of a vehicle beginning in state $(i, u)$ and taking action $a_j$ to park in a different lot $j$. If this succeeds, it incurs reward $r_{i, (j, o)} = -t_{i \rightarrow j} - t_{j \rightarrow D}$, where $t_{i \rightarrow j}$ represents the \textbf{drive time} from $i$ to $j$ and $t_{j \rightarrow D}$ represents the \textbf{walk time} from lot $j$ to the true destination $D$. If the attempt fails, it only incurs drive time reward $r_{i, (j, u)} = -t_{i \rightarrow j}$. If the vehicle remains unparked at any lot, it has two options for the following timestep: either seek to park at a new lot $k$ or else wait at the current lot for another chance at parking. The former incurs a reward of the formula just described, $r_{j, (k, u)}$ or $r_{j, (k, o)}$. Let $t_{\text{wait}}$ indicate a \textbf{wait time} incurred for remaining at a lot and waiting for another chance at parking. The latter incurs a reward $r_{j, (j, o)} = -t_{\text{wait}} - t_{j \rightarrow D}$ if it successfully parks and $r_{j, (j, u)} = -t_{\text{wait}}$ if it does not. In summary, the possible rewards can be described: 
\begin{equation*}
    r_{i, (j, \text{ status}=\{u, o\})}=\begin{cases}
    -t_{i \rightarrow j} - t_{j \rightarrow D}, & \text{if $i \neq j$ \& status $=o$}\\
    -t_{i \rightarrow j}, & \text{if $i \neq j$ \& status $=u$}\\
    -t_{\text{wait}} - t_{j \rightarrow D}, & \text{if $i = j$ \& status $=o$}\\
    -t_{\text{wait}}, & \text{if $i = j$ \& status $=u$}
    \end{cases}.
\end{equation*}
\ch{This reward structure allows us to formally distinguish between the conventional \textbf{time-to-drive} metric and our proposed \textbf{time-to-arrive} metric. We define the naive time-to-drive estimate as the simple travel time from origin to a target lot $j$, assuming immediate availability: $t_{0 \rightarrow j}$. In contrast, we define time-to-arrive as the magnitude of the cumulative reward collected over the entire trajectory $\tau$ until a terminal parked state is reached:$- \sum_{h \in \tau} r_h.$ Unlike time-to-drive, time-to-arrive captures the stochastic components of the journey, including potential multiple driving legs ($t_{i \rightarrow j}$) and dwell times ($t_{\text{wait}}$), as well as the final walking leg ($t_{j \rightarrow D}$).}

\ch{This distinction is critical for fair comparison across transportation modes. Current navigation platforms present time-to-drive estimates alongside time-to-arrive mass transit estimates that include wait and access/egress walk times. By omitting search and walking components, standard driving estimates present a structurally optimistic bias. Users, however, may compare app-displayed values across modes as equivalent estimates for total travel time. By explicitly modeling and optimizing for time-to-arrive, we correct this asymmetry, enabling valid cross-mode comparisons.}

The objective is \ch{therefore} to find the action sequence that \ch{minimizes expected time-to-arrive} (equivalently, \ch{maximizes the expected sum of rewards}). This framework excludes additional driver preferences, such as pricing; future work could integrate these. The MDP can be seen as a stochastic shortest path problem in which the terminating goal states $G \subset \mathcal{S}$ are the parked states; this termination condition obviates the need for a discount factor $\gamma$ in the framing.

\subsection{Optimal strategy and cost with probabilistic information}

This work focuses on the case in which probabilistic parking availability information is accessible for each lot since it most closely relates to today's app-based driving decisions. Given pre-existing awareness of parking difficulty and app-generated data, it is uncommon to truly have no prior information about parking availability odds~\cite{arora2019hard}. At the same time, real-time parking occupancy sensors are expensive and thus rare~\cite{djuric2016parkassistant}. 

Our analysis finds the optimal strategy and associated time cost falls into two structure-based regimes, described in turn by the propositions below.
\begin{prop}
    If, $\forall (i, j)$ pairs, $t_{i \rightarrow j} \geq t_{\text{wait}}$ and $t_{0 \rightarrow i} = t_{0 \rightarrow j}$, then the optimal strategy is to drive directly to the lot $i^*$ with the maximum value-to-go 
    \begin{equation}
        V_{i^*} = -t_{i^* \rightarrow D} - \frac{1}{p_{i^*}} t_{\text{wait}}.
    \end{equation}
\label{prop:patient}
\end{prop}

Note that the decision to park at lot $i$ can be conceptualized as a coin flip that succeeds (allows parking) with probability $p_i$. $t_{\text{wait}}$ can be considered the time one must wait at a lot to again `flip the coin' if the previous attempt was unsuccessful. Adopting a dynamic programming approach, one can thus explicitly write the cost for any `patient' strategy, i.e., the strategy in which a vehicle drives to the $i$th lot and waits there until it finds parking (possible since $p_i \in (0, 1]$). 
The expected cumulative return $\mathbf{E}[R_{i, \text{patient}}]$ of this strategy is written as
\begin{equation}
    \begin{split}
        \mathbf{E}[R_{i, \text{patient}}] = - & t_{0 \rightarrow i} - t_{i \rightarrow D} \\& - \sum_{1 \leq m \leq \infty} m \cdot t_{\text{wait}} \cdot (1 - p_i)^{m-1}p_i,
    \end{split}
\end{equation}
where the sum corresponds to the expected value of a geometric random variable with success probability $p_i$. We can thus rewrite the above as
\begin{equation}
        \mathbf{E}[R_{i, \text{patient}}] = -t_{0 \rightarrow i} - t_{i \rightarrow D} - \frac{1}{p_i} t_{\text{wait}}.
\label{eqn:expected_rwd}
\end{equation}
We can write the value-to-go of any unparked state $(i, u)$ under a patient policy at all times as 
\begin{equation}
V_{(i, u), \text{patient}} = - t_{i \rightarrow D} - \frac{1}{p_i} t_{\text{wait}}. 
\end{equation}
By the second assumption in the proposition statement (namely, $\forall (i,j)$ pairs, $t_{0 \rightarrow i} = t_{0 \rightarrow j}$), we have that
\begin{equation}
i^* = \mathop{\arg\!\max}\limits_{i \in \{ 0, \dots, N \} } V_{(i, u), \text{patient}} = \mathop{\arg\!\max}\limits_{i \in \{ 0, \dots, N \} } \mathbf{E}[R_{i, \text{patient}}]. 
\end{equation}
While this may seem a strong assumption, drive times to parking lots near a destination of interest are often quite comparable, especially relative to drive time variance due to other factors, \chold{e.g., traffic congestion, weather~\cite{peer2012prediction}. As an indication of this, empirical distributions of drive times on urban roads have been shown to exhibit standard deviations one-third of their means~\cite{fosgerau2012valuing}. Since we are concerned with \textit{a priori} destination selection, this variance enhances the extent to which drive times to two nearby locations may appear functionally similar, especially in comparison to the other time costs involved, thus expanding the proposition's scope}.


The question now becomes whether one can do better than a patient strategy, i.e., by switching to a new lot $j \neq i^*$ in the case where parking in a chosen lot $i^*$ is unsuccessful. However, note that this would incur the additional time cost of $t_{i^* \rightarrow j}$, which by assumption is at least as large as $t_{\text{wait}}$. 
By this fact and by the assumption that $t_{0\rightarrow i^*} = t_{0 \rightarrow j}$, we have 
\begin{equation}
    \begin{split}
        -t_{0 \rightarrow i^*} & - t_{\text{wait}}  + V_{i^*, \text{patient}}  \geq -t_{0 \rightarrow j} - t_{\text{wait}}  \\ & + V_{j, \text{patient}} 
         \geq -t_{0 \rightarrow i^*} - t_{i^* \rightarrow j} + V_{j, \text{patient}},
    \end{split}
\end{equation}
 where we also use the fact that $V_{i^*, \text{patient}} \geq V_{j, \text{patient}},\;\forall j$ due to the optimality of $i^*$. That is, any impatient strategy switching to a lot $j$ and terminating at lot $j$ will have a worse expected return than a patient strategy terminating at lot $i^*$. Therefore, in this regime, the optimal policy is a patient one in which the vehicle drives from the origin directly to lot $i^*$.  

\begin{prop}
    If $\exists t_{j \rightarrow k} < t_{\text{wait}}$, there may exist a cluster of parking lots $\{j, \dots, k\}$ such that it is better to visit the lots in that cluster rather than adopt a patient strategy at a parking lot $i$ with the single best value $V_{(i, u)}$.
\label{prop:cluster}
\end{prop}

Consider a cluster $\mathcal{C}$ of parking lots defined as those for which $t_{i \rightarrow j} < t_{\text{wait}}$ and $t_{j \rightarrow i} < t_{\text{wait}}$. Following from the above analysis, we can write the expected cumulative return for the strategy in which a vehicle navigates to this cluster and cycles through those lots until parking is found as 
\begin{equation}
    - t_{0 \rightarrow \mathcal{C}} - t_{\mathcal{C} \rightarrow D} + \frac{1}{1 - \prod_{i \in \mathcal{C}} (1 - p_i)}  \max\{-t_{\text{wait}}, -t_{\mathcal{C}}\},
\end{equation}
where $1 - \prod_{i \in \mathcal{C}} (1-p_i)$ is the probability that parking is available at any lot $i \in C$; 
$t_{\mathcal{C}}$ represents the time to navigate among lots in that cluster; 
the max operator reflects our modeling assumption of the $t_{\text{wait}}$ waiting time between two consecutive `coin flips' at the same lot; 
and $t_{0\rightarrow \mathcal{C}}$, $t_{\mathcal{C} \rightarrow D}$ respectively represent the travel time from the origin to the cluster and from the cluster to the destination (both can be bounded above by taking their maximums among lots within the cluster). 
There are settings in which cycling through a cluster $\mathcal{C}$ of parking lots is better than the optimal single-lot patient strategy on lot $i^*$, even when lot $i^*$ is not in the cluster $\mathcal{C}$. These settings correspond to scenarios when (1) the joint probability $1 - \prod_{i \in \mathcal{C}} (1-p_i)$ is high; (2) the travel time within the cluster $t_\mathcal{C}$ is small; and (3) the travel time $t_{0 \rightarrow \mathcal{C}} + t_{\mathcal{C}\rightarrow D}$ is low. Intuitively speaking, these criteria reflect the fact that after an unsuccessful parking attempt at a lot $i \in \mathcal{C}$, we can benefit from trying a different lot $j \in \mathcal{C}, j \neq i$ during the wait for the next `coin flip' at lot $i$, if we can travel from lot $i$ to $j$ relatively quickly, and lot $j$ has a relatively high probability of parking. If the cluster $\mathcal{C}$ additionally incurs relatively low travel times with respect to the origin and destination, the cluster strategy could be better than the single-lot optimal patient strategy, where a time period of $t_{\text{wait}}$ would be wasted between each `coin flip' at $i^*$. On the other hand, if these conditions are not satisfied, then the single-lot optimal patient strategy at $i^*$ would remain optimal in the probabilistic availability information regime.

\subsection{Sensitivity of the patient strategy}

\begin{prop}
    If, $\forall (i, j)$ pairs, $t_{i \rightarrow j} \geq t_{\text{wait}}$ and $t_{0 \rightarrow i} = t_{0 \rightarrow j}$, then the strategy of driving directly to lot $i^*$ with the maximum value-to-go $V_{i^*}$ remains optimal so long as, $\forall j \in N$, 
    \begin{equation}
        \left(\frac{1}{p_{j}} - \frac{1}{p_{i^*}}\right) \geq t_{0 \rightarrow i^*} - t_{0 \rightarrow j} + t_{i^* \rightarrow D} - t_{j \rightarrow D}.
    \end{equation}
\label{prop:sensitivity}
\end{prop}

This follows from the fact that the expected cumulative returns for $i^*$ and $j$ can be described as $\mathbf{E}[R_{i^*, \text{patient}}] \geq \mathbf{E}[R_{j, \text{patient}}]$. Rewriting with Proposition~\ref{prop:patient} and rearranging, one arrives at the result. Note the terms to the left of the ``+" sign on the right-hand side of the inequality represent the difference in drive times from the origin to lots $i^*$ and $j$, while the terms to the right of that same ``+" sign represent the difference in walk times from lots $i^*$ and $j$ to the destination. 

It is instructive to consider this in a decentralized, multi-agent context.
If each of the vehicles in this setting has an optimal strategy as described above, that strategy remains optimal as long as the conditions described in Proposition~\ref{prop:sensitivity} are maintained for that vehicle. Thus, the result is that no vehicle benefits by `defecting' to another parking lot and a system-wide Nash equilibrium is maintained. However, if the conditions are not maintained, then the equilibrium may shift.

\begin{figure*}
    \centering
    \captionsetup[subfloat]{font=scriptsize}
        \subfloat[First-order case]{\includegraphics[width=0.28\textwidth]{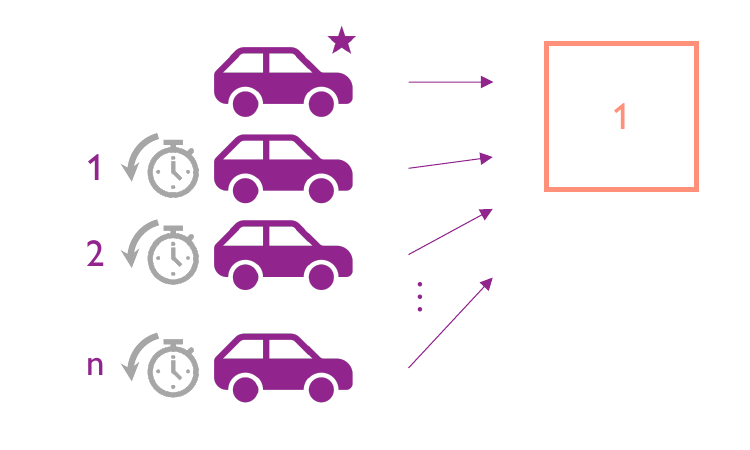}%
        \label{fig:case_1}}
        \hfil
        \subfloat[Second-order case]{\includegraphics[width=0.28\textwidth]{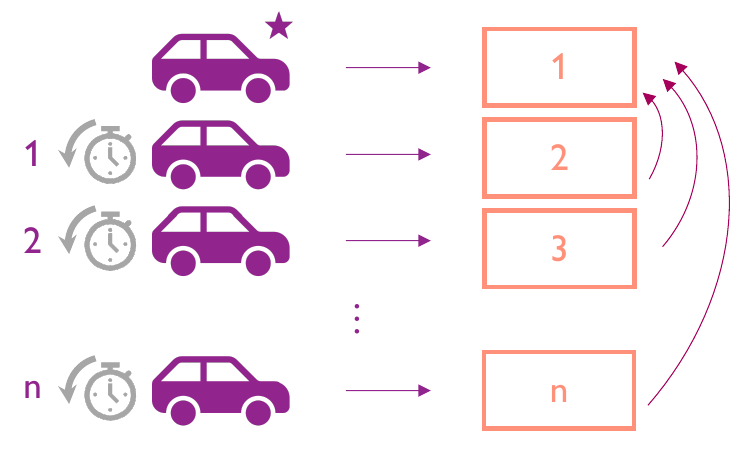}%
        \label{fig:case_2}}
        \hfil
        \subfloat[Third-order case]{\includegraphics[width=0.28\textwidth]{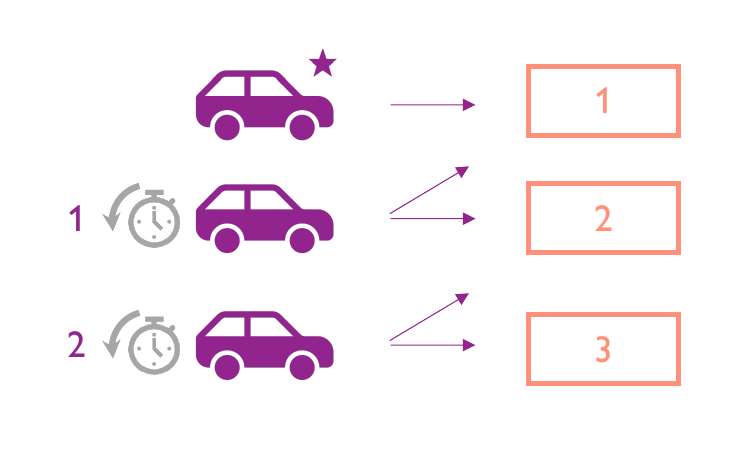}%
        \label{fig:case_3}}
    \caption{Visualization of the three illustrative cases for dynamic probabilities. The ego vehicle is indicated with a star, the rectangles represent parking lots, and vehicles are arranged such that lower vehicles arrive earlier.}
    \label{fig:illustrative_cases}
\end{figure*}

\subsection{Dynamic probabilities \& three illustrative cases}

Given the above sensitivity analysis, we now consider dynamic probabilities. In many settings, we do not expect walk time estimates between a parking lot and a destination of interest to vary considerably, and estimating driving time is a well-known feature of many of today's navigation apps as well as the subject of a number of previous works~\cite{bertsimas2019travel, uchida2014estimating, wang2018will}. Conversely, the probability of parking at a specified lot may vary considerably throughout the day due to demand, duration of individual stays at a location, etc. 

We investigate the impact of other vehicles in three settings that illustrate key mechanisms by which an ego vehicle's probability of parking can change. This investigation simultaneously exhibits the dynamic programming framework's ability to accommodate these analyses and to produce closed-form expressions for the resulting parking probabilities, which demonstrates how one could readily consider alternative situations. We consider three `orders' of impact, shown in Figure~\ref{fig:illustrative_cases} and described as follows:

\begin{itemize}
    \item \textbf{First-order}: Other vehicles attempt to park at the same lot as the ego vehicle. 
    \item \textbf{Second-order}: Other vehicles attempt to park at nearby lots, but resort to the ego vehicle's lot if parking at their first-choice lot is unavailable. 
    \item \textbf{Third-order}: Another vehicle does not attempt to park in the ego vehicle's lot, but affects the ego vehicle's parking probability via impacts on an intervening parking lot.  
\end{itemize}
The cases are not intended to be exhaustive; rather, they provide insight into the mechanisms by which a vehicle's parking probability may vary as well as the associated mathematics.

\subsubsection{First-order case -- arrivals to the same lot}

Assume that $n$ vehicles arrive to the same lot 1 as an ego vehicle within some near-term time $t < t_{\text{wait}}$ prior to that vehicle's arrival. The probability for that lot $p_1$ thus becomes $p_1'$, where $p_1' = p_1^n$; we can conceptualize it as multiple coin flips on the same `turn.' As the $n$th vehicle of this series to arrive, the ego vehicle can only park if all the vehicles prior were able to park and there is space remaining. 

We treat these as independent events in this analysis, which is reasonable in large parking lots with sufficient arrival and departure rates, when the time interval considered is appropriate, or when the true parking probability at a given lot is modeled as an underlying distribution. However, it may be a stronger assumption in other contexts. At first, it seems to fail to take advantage of additional information, namely, any vehicle which successfully parks prior to the ego vehicle ought to lower our estimate of the ego vehicle's ability to park. However, a successful parking event provides information which can counteract this.
For example, if a vehicle successfully parks, it may indicate timing is favorable or that more vehicles are entering than exiting. Thus, it may not be appropriate to reduce a parking probability estimate for the ego vehicle conditional on a prior vehicle's successful parking event. Additionally, the analysis and experiments further below consider the error incurred when we have access to only intermittent observations of a dynamic probability, such as can occur if arriving vehicles do affect the underlying probability. 

One interesting observation is that in the settings this work considers, the online setting is nearly equivalent to the offline setting, and indeed entirely equivalent if we allow the online setting some `lookahead' time. This is because other vehicles arriving to a lot after a vehicle is parked do not affect that vehicle's odds of parking, and in the decentralized setting with self-interested agents there is no incentive to make suboptimal decisions in the short term to enhance parking options for vehicles yet to come. Furthermore, the lookahead time need only be as long as the time between the ego vehicle's departure and its parking, since the gap between useful online and offline knowledge is simply the arrival of other vehicles to lots of interest in that period. This could be possible to some extent in settings where users were navigating to parking lots based on the recommendations from a navigation app. This lookahead time also presents a natural choice for the near-term time $t < t_{\text{wait}}$ mentioned above. 

\subsubsection{Second-order case -- arrivals to nearby lots}

Consider when near-term prior arrivals occur at lots near the target lot for the ego vehicle (lot 1), but only proceed to that lot if parking at their `first-choice' lot fails. This could occur for a variety of reasons, including if vehicles are pursuing the cluster-based strategy mentioned in Proposition~\ref{prop:cluster}. In this case, the updated probability for lot 1 becomes 
\begin{equation}
    p_1' = \prod_{j=1}^{n}p_j + \sum_{k=2}^{n}p_1^k (1 - p_k) \prod_{l=k+1}^{n}p_l.
\end{equation}

The proof follows from the fact that we can conceptualize of this as a series of coin flips, where the situation in which other vehicles attempt to park at lot 1 is a result of vehicles' failures to park in other lots, and thus subject to those probabilities. One note: in the final element of $\prod_{l=k+1}^{n}p_l$, $k=n$ would mean that $l > n$, which would be beyond the allowable bounds for $l$. Thus, in this case, we simply set that term to 1. 

\subsubsection{Third-order case -- knock-on effects}

Whereas Cases 1 and 2 consider $n$ vehicles beyond the ego vehicle, for ease of exposition Case 3 only considers two other vehicles. More specifically, a third vehicle attempts to park at lot 3 prior to the others, but if it fails it will attempt to park at lot 2. That affects a second vehicle's ability to park at lot 2, which is its first-choice lot. If the second vehicle cannot park at lot 2, it will attempt to park at the ego vehicle's desired lot. In this way, the third vehicle affects the ego vehicle's ability to park despite never attempting to park in its lot. The proof follows directly from this logic. The updated probability $p_1'$ for the ego vehicle can be written as 
\begin{equation}
    \begin{split}
        p_1' = p_3 (p_2 p_1 + & (1 - p_2) p_1^2) + (1 - p_3) [ p_1 p_2^2 + \\&p_1^2 (1 - p_2^2 - (1 - p_2)^2) + (1 - p_2)^2 p_1^3  ].
    \end{split}
\end{equation}
By way of interpretation, the $p_1 p_2^2$ term indicates the case in which both the second and third vehicles successfully park in lot 2. The $p_1^2$ indicates the case in which \textit{either} the second \textit{or} the third vehicle successfully parks in lot 2. The $(1 - p_2)^2$ term indicates the case in which neither of those vehicles is successful in parking in lot 2, and thus proceed to lot 1. 

\subsection{Errors from \chold{stochastic probability observations}}

\begin{figure}
    \centering
    \captionsetup[subfloat]{font=scriptsize}
        \subfloat[Linear increase]{\includegraphics[width=0.21\textwidth]{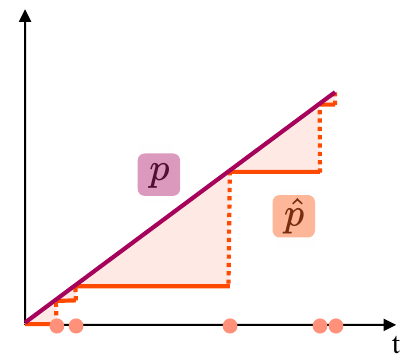}%
        \label{fig:linear_probs_up}}
        \hfil
        \subfloat[Linear decrease]{\includegraphics[width=0.21\textwidth]{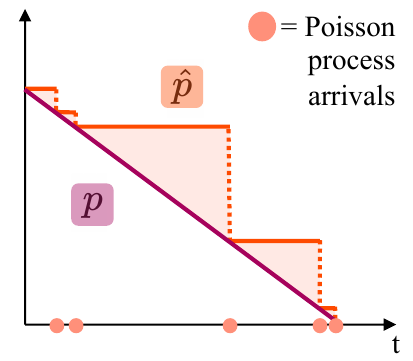}%
        \label{fig:linear_probs_down}}
    \caption{Error (shaded area) between an estimate $\hat{p}$ and the true value $p$ when $p$ is varying linearly and observation times are generated via a Poisson process.}
    \label{fig:linear_probs}
\end{figure}

If the underlying lot-level probabilities are dynamic, \chold{stochastic measurements} will incur error as their recency fades. This can be an issue in light of the sensitivity analysis above; i.e., if the observed probabilities deviate too much from the true probabilities, vehicles might route to suboptimal locations. Thus, we provide analysis and experiments below to investigate errors incurred when estimating dynamic probabilities via Poisson process-generated observations.
Poisson processes are a well-known stochastic modeling tool for random arrival processes, \chold{ including for parking analysis \cite{newell1982,Jorge2012,kutoyants2023, xiao2018likely, ogulenko2022nature}.}

\chold{The analysis below applies to any pointwise-independent stochastic observation method and could easily be adapted to a variety of schemes.} \cite{xiao2023parking} provides a comprehensive survey on parking prediction, particularly via data-driven methods for probability prediction. Near-term prediction is also discussed in~\cite{xiao2018likely}. \chold{For concreteness -- and given the costs and effort associated with permanent infrastructure, continuous remote sensing, etc. -- the current work considers that these Poisson} observations are generated by connected users, where such users represent $r$ fraction of the total population. \chold{This is inspired by the approach in~\cite{shi2018parkcrowd}.} For example, users of a given routing app or connected vehicles (autonomous or not) could provide insight into the availability at a lot to which they arrive or pass. \cite{bock2019smart} has illustrated the feasibility of a similar method, although that work leverages taxis and considers only on-street parking. \chold{Note, however, that it would be straightforward to adapt the analysis to other Poisson stochastic observation methods, including schemes with remote sensing or paid staff, by adjusting the rate parameter.}

\begin{prop}
    With vehicles arriving according to a Poisson process parameterized by arrival rate $\lambda$ and connective technology adoption rate $r$, the expected (mean) error between the true value of a variable $p$ varying in a linear fashion with slope $m$ and observations from connected users of that variable $\hat{p}$ will be ${m}/{\lambda^2 r^2}$.
\label{prop:linear_error}
\end{prop}

This can be found by taking the expectation over the integral of a linear function with slope $m$ (e.g., $y=mx$) with respect to the integral's upper bound, where the upper bound is a random variable generated via a Poisson process with rate parameter $\lambda r$. See Figure~\ref{fig:linear_probs} for a visualization of the Poisson process-guided observation process, as well as the errors incurred. The result in Proposition~\ref{prop:linear_error} corresponds to the average size of the shaded regions in Figure~\ref{fig:linear_probs}. Each observation of $\hat{p}$ is held constant until a new one is made. 

\begin{prop}
    With vehicles arriving according to a Poisson process parameterized by arrival rate $\lambda$ and connective technology adoption rate $r$, the expected (mean) error between the true value of a variable $p$ increasing in an exponential fashion with exponent $b$ and observations from connected users of that variable $\hat{p}$ will be ${b}/{(\lambda r)^{b+1}}$.
\end{prop}

As before, this can be found by taking the expectation over the integral of an exponential function with exponent $b$ (e.g., $y=x^b$) with respect to the integral's upper bound, where the upper bound is a random variable generated via a Poisson process with rate parameter $\lambda r$. Unlike before, the proof leverages the definition of the moments of an exponential distribution. This would correspond to the average size of the shaded regions in Figure~\ref{fig:linear_probs} if the true probability function $p$ were exponentially increasing. 

\section{\chold{Numerical Experiment: Stochastic Observation Error}}
\label{sec:experiments_stoch}

\begin{figure}
    \centering
    \captionsetup[subfloat]{font=scriptsize}
    \subfloat[]{\includegraphics[width=0.25\textwidth, clip=true, trim=0 10 0 10]{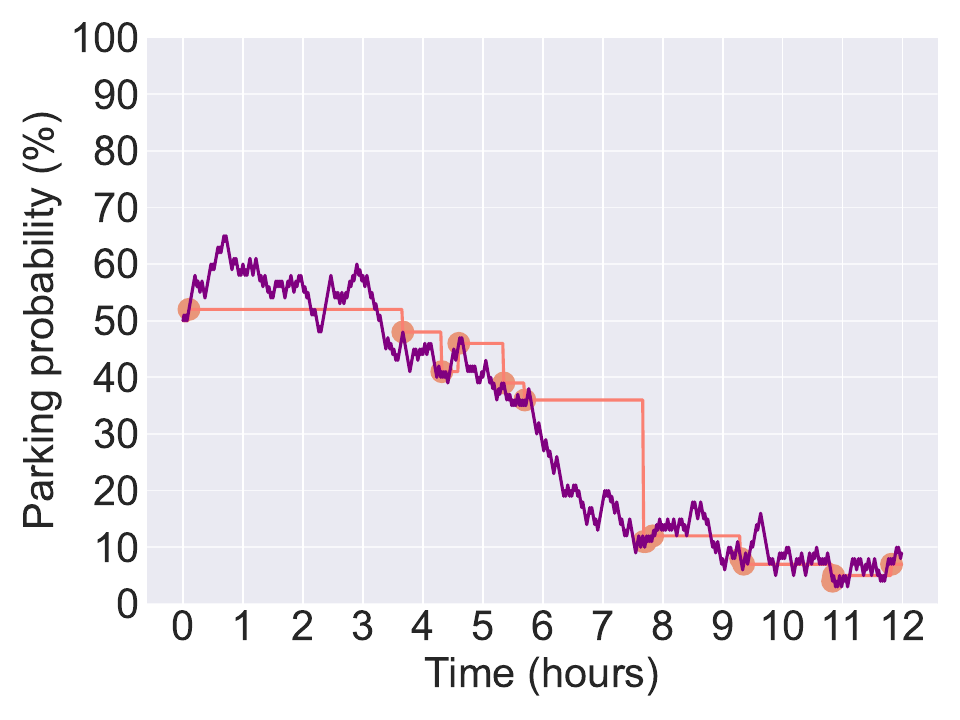}
    \label{fig:random_walk_low}}
    \subfloat[]{\includegraphics[width=0.207\textwidth, clip=true, trim=80 10 0 10]{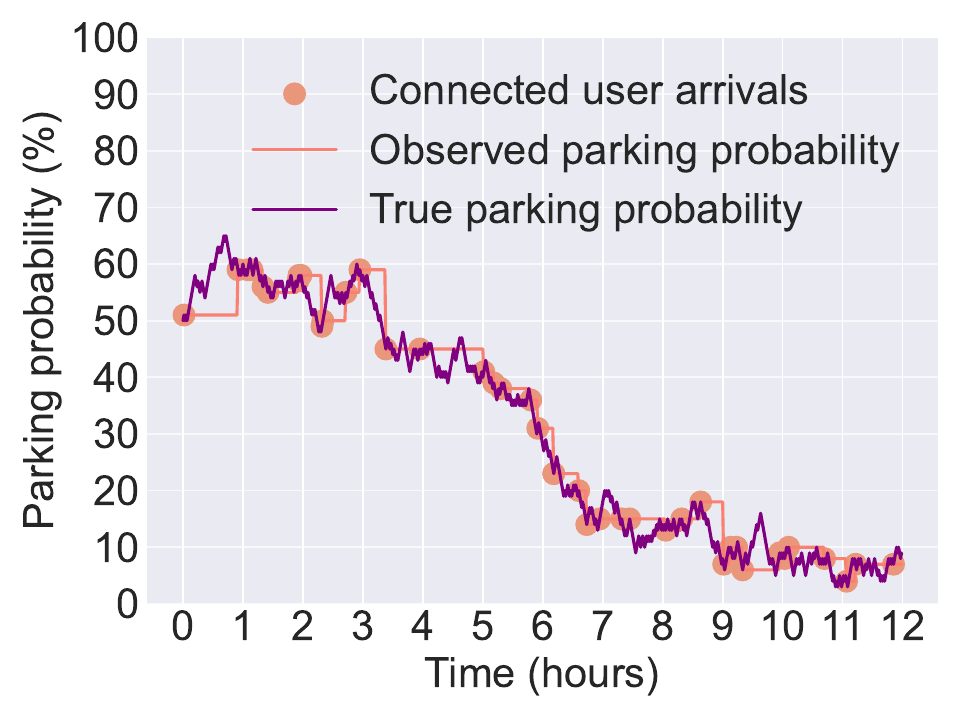} 
    \label{fig:random_walk_high}}
    \caption{Discrepancy in observed and true probabilities when the true probability is a bounded random walk beginning at 50\% for (a) lower observation rate ($\lambda=10$ veh/hr, $r=10$\% adoption) and (b) higher observation rate ($\lambda=20$ veh/hr, $r=20$\% adoption). The observation rate is a function of both the overall vehicle arrival rate $\lambda$ and connected user adoption fraction $r$.}
    \label{fig:random_walks}
\end{figure}
\begin{figure} [t]
    \centering
    \captionsetup[subfloat]{font=scriptsize}
    \subfloat[]{\includegraphics[width=0.25\textwidth, clip=true, trim=0 10 0 10]{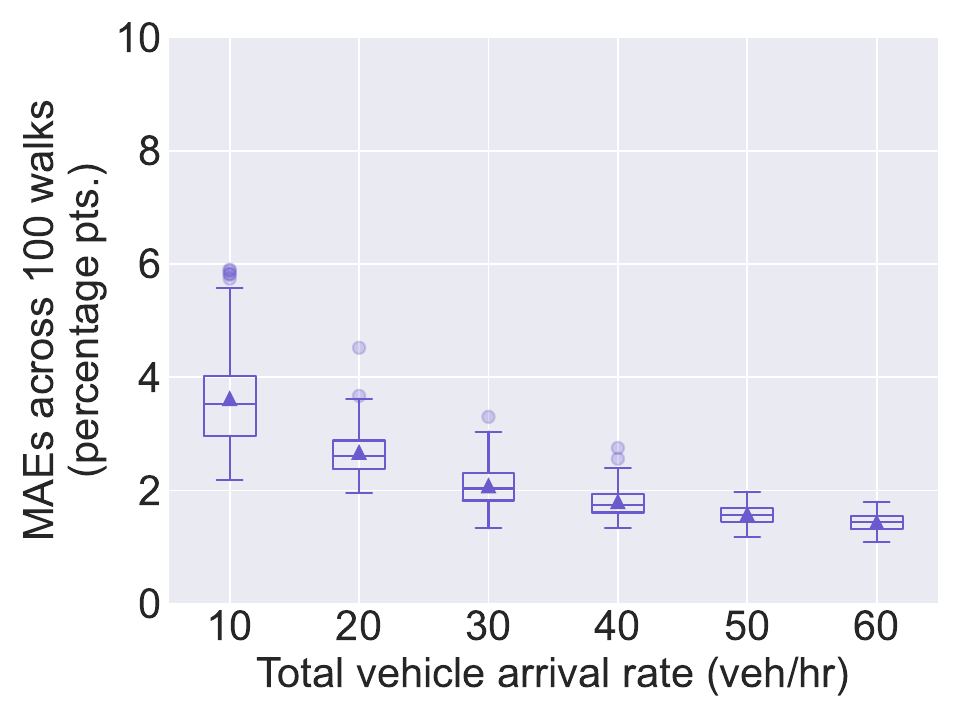}%
    \label{fig:error_curve_arrival}}
    \subfloat[]{\includegraphics[width=0.207\textwidth, clip=true, trim=80 10 0 10]{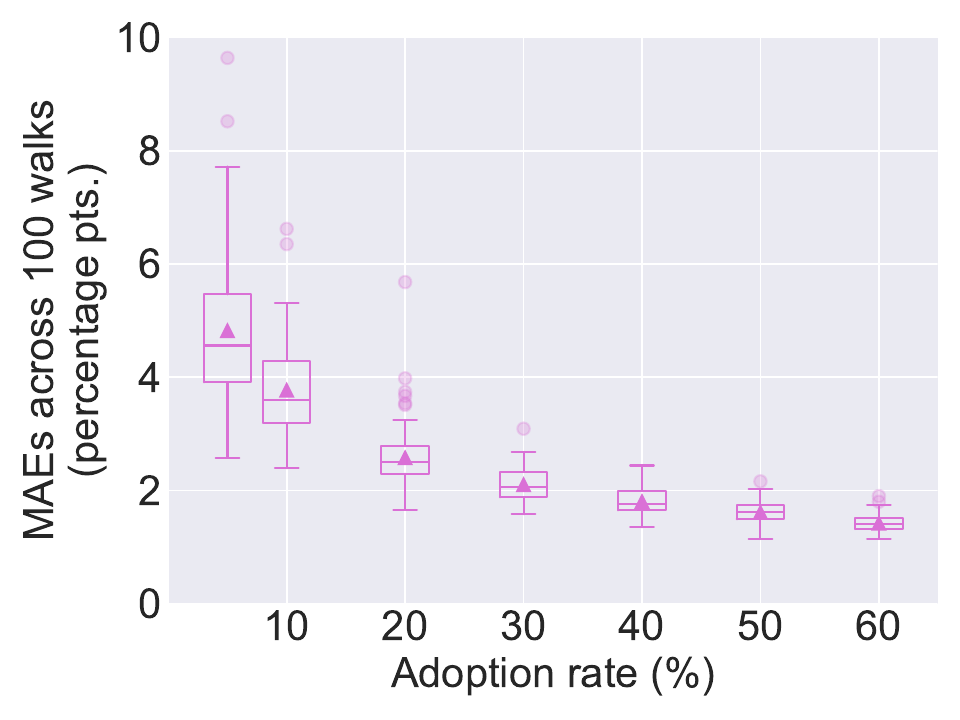}
    \label{fig:error_curve_adoption}}
    \caption{\chold{Distribution of mean absolute errors (MAEs) between observed and true probabilities across 100 different random walks as a function of (a) arrival rates (with $r=20$\%) and (b) adoption rates (with $\lambda=20$). The distributional mean is indicated by a triangle and individual MAEs more than 150\% beyond each box's inter-quartile range are indicated with dots.}}
    \label{fig:error_curves}
\end{figure}

The probabilities in the above analysis increased or decreased monotonically. This may be relevant in some settings, such as in the lead-up to or aftermath of a major event, or for commuter parking lots at the beginning and end of a workday. Obviously, these fail to characterize parking lots for which the dynamics are less predictable. 

To accommodate this, we evaluate via simulation the errors incurred when making intermittent (Poisson process-generated) observations of a bounded random walk. The walk is bounded between 0\% and 100\% to mirror true probabilities and increments or decrements by one percentage point each minute. As before, observations are generated via a Poisson process with rate $\lambda r$. Observations are again retained until a new one is made. Figure~\ref{fig:random_walks} provides a visualization of one such random walk under two separate observation rates. 

    To get a better sense of error, we also assess average error as a function of $\lambda$ and $r$. Results are shown in Figure~\ref{fig:error_curves}. These were generated by simulating the bounded random walks across \chold{100} seeds at each arrival or adoption rate. All began at a probability value of 50\% and run for 12 hours of simulated time. Not only does the error decrease with more users, but in all settings studied, it remains relatively low, with mean absolute errors averaging below 5\%. For reference, with a drive, walk, and wait time of 15, 10, and 5 minutes respectively, a 5\% overestimate of a 45\% parking probability at a lot amounts to an expected travel time error in Equation \ref{eqn:expected_rwd} of less than one minute. Code and data for this paper are available at \texttt{\url{https://github.com/chickert/Probability-Aware-Parking-Selection}}.

\section{\chold{Case study: Seattle parking}}
\label{sec:seattle_sim}

\begin{figure} [t]
	\center
	\includegraphics[width=0.233\textwidth, clip=true, trim=0 25 0 0]{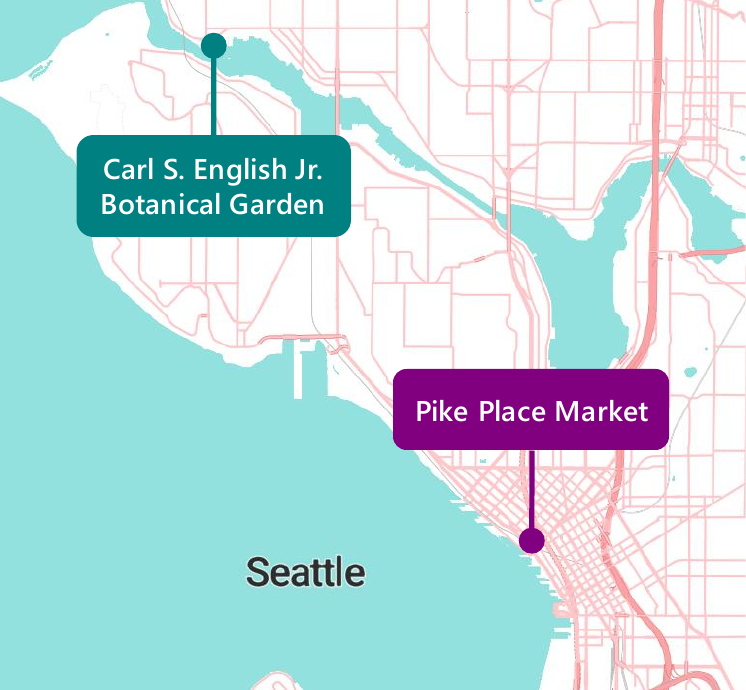}
	\caption{\ch{Locations of the destinations used in the Seattle case study. Base map sources: MapTiler, OpenStreetMap contributors.}}
\label{fig:seattle}
\end{figure}

\subsection{\chold{Probability observation errors}}
\label{subsec:prob_estimation}

\subsubsection{{\chold{Sites}}}
To validate \chold{the results of the proposed estimation approach across different times of day and urban parking settings}, we introduce data provided by Seattle’s Department of Transportation (SDOT)~\cite{seattle_paid_parking_occupancy, seattle_paid_parking_transactions} for two separate areas: 
\chold{\begin{itemize}
    \item (Dense) \textbf{Area A.1}: 61 SDOT paid parking spots around the Pike Place Market (47.60943, -122.34183).
    \item (Sparse) \textbf{Area B.1}: 72 spots around the Carl S. English Jr. Botanical Garden (47.66687, -122.39747). 
\end{itemize}}
\noindent\chold{They are selected to reflect heterogeneity:} Area A.1 is a dense urban setting in the heart of downtown Seattle with high traffic, while Area B.1 is located five miles northwest and represents an area with lighter traffic, \chold{less dense zoning, and sparser SDOT parking options}.
\ch{See Figure~\ref{fig:seattle}.}

\subsubsection{\chold{Scenario}}
We use payment transactions collected on Thurs., Jan. 30th, 2025 as evidence of arrivals and take inverse occupancy rates as a proxy for parking probability. \chold{The data is available for Area A.1 from 8am to 8pm and for Area B.1 from 8am to 6pm.} Occupancy is computed from paid parking occupancy and the total parking space count. To simulate connected users, we randomly sample the transaction data and assign the appropriate proportion as connected user arrivals.

\subsubsection{Results}
\begin{figure} [t]
    \centering
    \captionsetup[subfloat]{font=scriptsize}
        \subfloat[]{\includegraphics[width=0.24\textwidth, clip=true, trim=0 10 0 10]{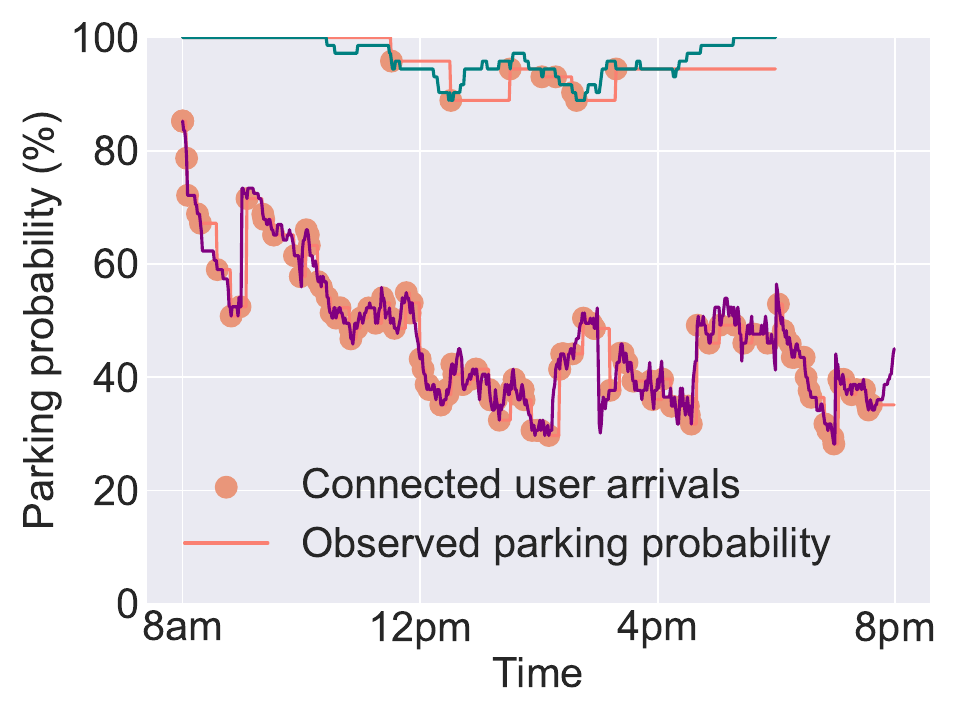}
        \label{fig:seattle_walk}}
        \subfloat[]{\includegraphics[width=0.24\textwidth, clip=true, trim=0 10 0 10]{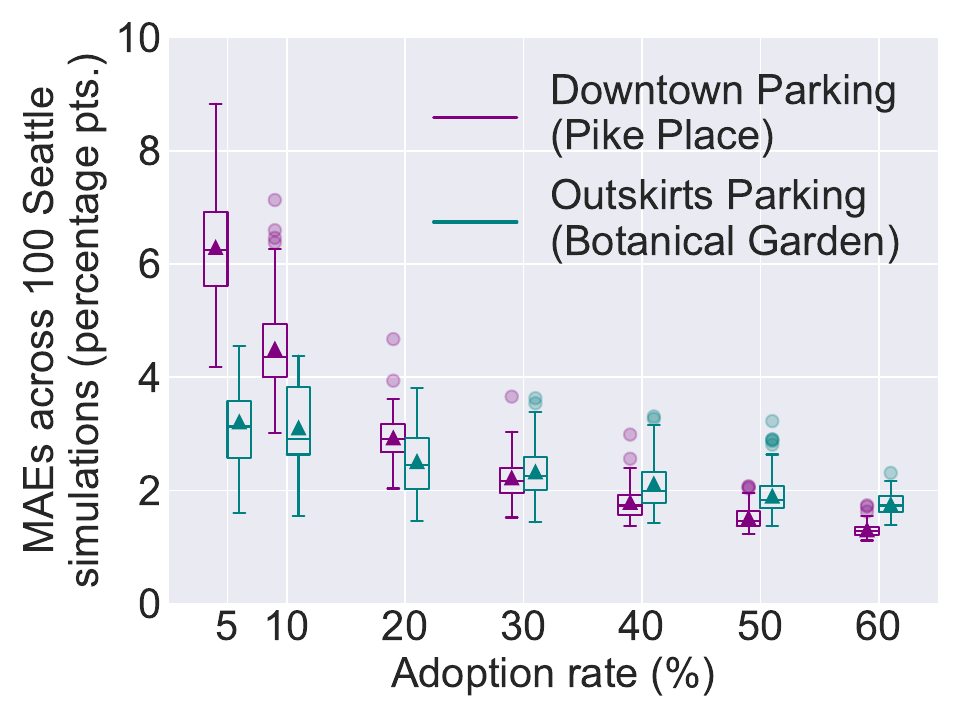}%
        \label{fig:seattle_error}}
        \caption{Dynamic probability results for two locations in Seattle: Pike Place Market (A.1, purple) and the Carl S. English Jr. Botanical Garden (B.1, teal). (a) Discrepancy in true and observed (orange) probabilities given the arrivals in SDOT data and a simulated $30$\% connected user adoption rate. (b) MAE distributions as the adoption rate increases. Plot conventions are as before.}
    \label{fig:seattle_plots}
\end{figure}

\chold{We evaluate stochastic probability observation via connected users by computing errors as in Section~\ref{sec:experiments_stoch}. Results are shown in Figure~\ref{fig:seattle_plots} with the same conventions as in Figures~\ref{fig:random_walks} and~\ref{fig:error_curves}. 
Given the connected user sampling involved, for the error box plots we again evaluated 100 random seeds at each adoption rate. In all cases the mean and median MAE is below 7\% and declines with increased adoption to below 2\%. Interestingly, while Figure~\ref{fig:seattle_walk} shows the relative discrepancy between true and observed probabilities in Area B.1 to be greater than that at Area A.1 due to the lower effective arrival rate of connected users, Figure~\ref{fig:seattle_error} indicates that -- by the same token -- the lower resulting volatility in Area B.1's parking probability produces MAEs that are more consistently low.
}

\begin{table*}
\caption{\chold{Comparison of Probability-Aware Selection Policies with Patient and Impatient Baselines Across Temporal and Spatial Settings}}
\label{tab:policy_performance}
\begin{tabular*}{\textwidth}{@{\extracolsep{\fill}}lllrrrrrrrr}
\toprule
 &  & & \multicolumn{4}{c}{10\% Adoption Rate} & \multicolumn{4}{c}{50\% Adoption Rate} \\
\cmidrule(lr){4-7} \cmidrule(lr){8-11}
Destination & Day & Policy & \makecell{Mean Time\\± Std. (\textdownarrow)} & \makecell{Gain vs.\\BL-Pat. (\textuparrow)}  & \makecell{Gain vs.\\BL-Imp. (\textuparrow)} & \makecell{Perf. vs.\\Oracle (\textuparrow)} & \makecell{Mean Time\\± Std. (\textdownarrow)} & \makecell{Gain vs.\\BL-Pat. (\textuparrow)}  & \makecell{Gain vs.\\BL-Imp. (\textuparrow)} & \makecell{Perf. vs.\\Oracle (\textuparrow)} \\
\midrule
\multirow[t]{10}{*}{\makecell{\\\\Area A.2\\(Pike Place\\Market)}} & \multirow[t]{5}{*}{\makecell{\\Week-\\day}} & BL-Patient & 44.6 ± 20.1 & --- & --- & --- & 45.2 ± 19.4 & --- & --- & --- \\
 &  & BL-Impat. & 21.9 ± 5.6 & \cellcolor{greendark}50.9\% & --- & --- & 20.8 ± 6.3 & \cellcolor{greendark}54.0\% & --- & --- \\
 &  & PA 1-Step & 20.0 ± 7.4 & \cellcolor{greendark}55.2\% & \cellcolor{greenlight}8.7\% & \cellcolor{redlight}-8.2\% & 19.6 ± 4.1 & \cellcolor{greendark}56.8\% & \cellcolor{greenlight}6.1\% & \cellcolor{redlight}-1.3\% \\
 &  & PA 2-Step & 19.6 ± 4.7 & \cellcolor{greendark}55.9\% & \cellcolor{greenlight}10.2\% & \cellcolor{redlight}-1.8\% & 21.4 ± 6.8 & \cellcolor{greendark}52.6\% & \cellcolor{redlight}-2.9\% & \cellcolor{greenlight}2.1\% \\
 &  & \textbf{PA 3-Step} & \textbf{19.0 ± 4.3} & \cellcolor{greendark}\textbf{57.4\%} & \cellcolor{greenlight}\textbf{13.2\%} & \cellcolor{greenlight}0.5\% & \textbf{19.1 ± 5.2} & \cellcolor{greendark}\textbf{57.7\%} & \cellcolor{greenlight}\textbf{8.1\%} & \cellcolor{redlight}-7.2\% \vspace{1mm}\\
 & \multirow[t]{5}{*}{\makecell{\\Week-\\end}} & BL-Patient & 55.6 ± 13.2 & --- & --- & --- & 56.2 ± 12.4 & --- & --- & --- \\
 &  & BL-Impat. & 27.3 ± 9.9 & \cellcolor{greendark}50.9\% & --- & --- & 25.3 ± 7.9 & \cellcolor{greendark}55.1\% & --- & --- \\
 &  & \textbf{PA 1-Step} & \textbf{21.9 ± 9.1} & \cellcolor{greendark}\textbf{60.7\%} & \cellcolor{greenmed}\textbf{20.0\%} & \cellcolor{redlight}-5.4\% & 19.7 ± 4.2 & \cellcolor{greendark}65.0\% & \cellcolor{greenmed}22.1\% & \cellcolor{greenlight}5.0\% \\
 &  & \textbf{PA 2-Step} & 22.3 ± 9.4 & \cellcolor{greendark}59.9\% & \cellcolor{greenlight}18.4\% & \cellcolor{redlight}-15.3\% & \textbf{19.1 ± 3.8} & \cellcolor{greendark}\textbf{66.1\%} & \cellcolor{greenmed}\textbf{24.5\%} & \cellcolor{redlight}-3.4\% \\
 &  & PA 3-Step & 22.4 ± 9.0 & \cellcolor{greendark}59.7\% & \cellcolor{greenlight}18.1\% & \cellcolor{redlight}-12.6\% & 19.8 ± 4.6 & \cellcolor{greendark}64.8\% & \cellcolor{greenmed}21.7\% & \cellcolor{greenlight}11.9\% \vspace{1mm}\\
\multirow[t]{10}{*}{\makecell{\\\\Area B.2\\(Botanical\\Garden)}} & \multirow[t]{5}{*}{\makecell{\\Week-\\day}} & BL-Patient & 10.3 ± 1.2 & --- & --- & --- & 10.2 ± 1.0 & --- & --- & --- \\
 &  & BL-Impat. & 10.8 ± 2.2 & \cellcolor{redlight}-4.4\% & --- & --- & 10.5 ± 1.9 & \cellcolor{redlight}-2.8\% & --- & --- \\
 &  & \textbf{PA 1-Step} & \textbf{10.0 ± 0.0} & \cellcolor{greenlight}\textbf{3.2\%} & \cellcolor{greenlight}\textbf{7.2\%} & \cellcolor{greenlight}1.1\% & \textbf{10.1 ± 0.7} & \cellcolor{greenlight}\textbf{1.1\%} & \cellcolor{greenlight}\textbf{3.8\%} & \cellcolor{greenlight}1.1\% \\
 &  & \textbf{PA 2-Step} & 10.8 ± 3.1 & \cellcolor{redlight}-4.4\% & 0.0\% & \cellcolor{redlight}-4.4\% & \textbf{10.1 ± 0.7} & \cellcolor{greenlight}\textbf{1.1\%} & \cellcolor{greenlight}\textbf{3.8\%} & \cellcolor{greenlight}1.1\% \\
 &  & \textbf{PA 3-Step} & 10.2 ± 1.0 & \cellcolor{greenlight}1.1\% & \cellcolor{greenlight}5.2\% & 0.0\% & \textbf{10.1 ± 0.7} & \cellcolor{greenlight}\textbf{1.1\%} & \cellcolor{greenlight}\textbf{3.8\%} & \cellcolor{redlight}-1.1\% \vspace{1mm}\\
 & \multirow[t]{5}{*}{\makecell{\\Week-\\end}} & BL-Patient & 14.6 ± 8.4 & --- & --- & --- & 14.0 ± 8.9 & --- & --- & --- \\
 &  & BL-Impat. & 13.8 ± 5.2 & \cellcolor{greenlight}5.1\% & --- & --- & 14.0 ± 5.5 & \cellcolor{redlight}-0.1\% & --- & --- \\
 &  & \textbf{PA 1-Step} & \textbf{13.4 ± 4.5} & \cellcolor{greenlight}\textbf{8.1\%} & \cellcolor{greenlight}\textbf{3.2\%} & \cellcolor{greenlight}1.1\% & \textbf{11.6 ± 2.4} & \cellcolor{greenlight}\textbf{17.1\%} & \cellcolor{greenlight}\textbf{17.3\%} & \cellcolor{greenlight}5.5\% \\
 &  & PA 2-Step & 13.5 ± 3.6 & \cellcolor{greenlight}7.3\% & \cellcolor{greenlight}2.4\% & \cellcolor{greenlight}1.3\% & 13.4 ± 3.4 & \cellcolor{greenlight}4.4\% & \cellcolor{greenlight}4.6\% & \cellcolor{redlight}-5.1\% \\
 &  & PA 3-Step & 13.8 ± 5.5 & \cellcolor{greenlight}5.1\% & 0.0\% & \cellcolor{redlight}-8.3\% & 13.6 ± 5.4 & \cellcolor{greenlight}3.0\% & \cellcolor{greenlight}3.1\% & \cellcolor{redlight}-9.7\% \\
\bottomrule
\vspace{0.1mm}
\end{tabular*}
\chold{For brevity, `PA' indicates probability-aware policies and `BL' indicates baselines. Up arrows indicate that higher values are preferable for the metric, while down arrows indicate the opposite. Times are reported in minutes. Boldface denotes the top performance within each adoption rate category. See the main text for further details. \ch{Darker shading indicates larger performance differences.}}
\end{table*}

\subsection{\chold{Strategy comparison}}
\label{subsec:strat_comparison}

\chold{To validate the relative merits of probability-aware parking selection with stochastic observations, we conduct comparisons against traditional (baseline) methods. This uses simulations of parking trials.}

\subsubsection{\chold{Sites}}
\chold{In this experiment, for each destination we now construct an MDP according to the framework in Section~\ref{sec:methodology}, with $N=3$ parking lots each.}
\begin{itemize}
    \item \chold{(Dense) \textbf{Area A.2}: Similar to Area A.1., Pike Place Market is the destination. The three lots, which comprise 98 spots, are those nearest the destination along Pike Place (Lot 1, closest), 2nd Avenue (Lot 2, next closest), and Alaskan Way (Lot 3, furthest). The origin is Kerry Park, a residential neighborhood chosen for its centrality within the city. Google Maps estimates an uncongested time-to-drive of ten minutes and public transit total travel times (time-to-arrive) of 18-22 minutes, depending on the time of day.}
    \item \chold{(Sparse) \textbf{Area B.2}: The Botanical Garden is the destination. The lots, which comprise 69 spots, include one on 54th St. (Lot 1) and the two nearest the Garden on Market St. (Lots 2 and 3, respectively). Again, Lot 1 is the nearest to the destination and Lot 3 is furthest. The origin is Salmon Bay Park, selected to provide geographic heterogeneity as a residential hub in outer Seattle. Google Maps estimates an uncongested time-to-drive of six minutes and public transit times of 21-23 minutes (again including walk time).}
\end{itemize}

All parking lots were selected as viable options (i.e., within a 10-minute walk from their destination) with differing parking availabilities. Origins are picked such that the drive times to each lot are approximately equal.

\subsubsection{\chold{Scenarios}}

\chold{To capture temporal heterogeneity, we simulate multiple days and departure times. The weekday selected is Thurs., Jan. 30th, 2025, and the weekend day selected is Sat., Feb. 1st, 2025. The selection of this weekend date for Area A.2 is intentional to consider event interference: a festival brought crowds to the area for music, art, and shopping.
}

\chold{For each day we simulate trips beginning each hour from 8am to 6pm for Area A.2 and from 8am to 4pm for Area B.2 (since SDOT parking data collection stops earlier for these lots). Connected user arrivals and observations are modeled as in Section~\ref{subsec:prob_estimation}. Results are reported for 10\% and 50\% user adoption rates to explore the impact of errors on system reliability, but were similar for other values. For robustness given the stochastic nature of the simulations, we run each simulation for five random seeds at each departure time across the locations, dates, and adoption rates. Drive times and walk times are estimates from Google Maps.  We set $t_{\text{wait}} = 5$ minutes. The true parking probabilities at each of the six lots are extracted from SDOT data as described in Section~\ref{subsec:prob_estimation}.
}

\subsubsection{\chold{Policies}} 
\chold{We design three probability-aware (PA) parking selection policies and two baseline methods against which to validate their relative merits. Dynamic probabilities preclude the possibility of computing the true value of each possible action. 
We do not assume future probabilities at each lot are known. The evaluated policies are:
\begin{itemize}
    \item \textbf{PA 1-Step} computes a probability risk-adjusted time cost heuristic $c_{i,j} = (t_{i \rightarrow j} / p_j) + t_{j \rightarrow D}$ for each lot $j$ from the current lot $i$ (including the possibility that $i = j$), then attempts to park in the lot with the lowest projected cost. Note this is a `one-step' heuristic in that it only considers the odds of parking at the target lot. 
    \item \textbf{PA 2-Step} `looks ahead' to consider the option(s) available should the immediate parking attempt fail. It is thus recursive. The two-step cost heuristic may be written $c_{i,j} = t_{i \rightarrow j} + p_j(t_{j \rightarrow D}) + (1 - p_j)(\min_{i \in \{0, ..., N\}}c_{j,i})$; note this last term is the cost heuristic for the best possible action if the first parking attempt fails.
    \item \textbf{PA 3-Step} is a three-step heuristic that extends the consideration one step further and follows the same mathematical structure.
\end{itemize}}

\chold{Note that we still do not assume access to future probability knowledge. Thus, while each additional step incorporates more knowledge about the system's current state, this may introduce error in a dynamic setting such as ours.}

\chold{Greedy patient and impatient policies serve as  baselines. These are derived from the recommendations of popular navigation apps and how users may behave in response, either by waiting or seeking additional nearby parking~\cite{arora2019hard}.} 
\chold{\begin{itemize}
    \item \textbf{Baseline-Patient} is a patient strategy that finds the lot closest to the destination and waits there until parked.
    \item \textbf{Baseline-Impatient} is an impatient policy that first attempts to park at the lot closest to the destination. If this fails, it drives to the unvisited lot closest to its current position to continue its search. If all lots have been visited, the search cycle resets.
\end{itemize}}
 
\chold{Since the patient policy may produce unrealistically long waits during low-availability times, we cap this policy's search at 1 hour to prevent skewed results.}

\subsubsection{\chold{Results}} 
\chold{For each simulation we compute the time-to-arrive. This is the per-episode cumulative reward, as described in Section~\ref{subsec:framework}. The results from these experiments are shown in terms of time savings and loss in Table~\ref{tab:policy_performance}. The means and standard deviations (in minutes) of trip durations are presented for each setting. Gains are computed as percent improvements over the baseline policies' mean times. Policies in bold indicate a top performance in \textit{either} the 10\% \textit{or} the 50\% adoption rate category. To better understand the causes of errors and their impacts on the system, we also compare probability-aware policies against oracle baselines that have access to the current true parking probabilities at each lot, rather than those generated from connected user observations. These quantify the effect of suboptimal probability awareness.}

\chold{The results indicate probability-aware parking selection strategies, even with probability estimates generated stochastically via connected users, outperform more traditional methods. Improvements reach up to 66\% relative to the patient baseline and 24\% relative to the impatient baseline. As hypothesized, these gains are greater when parking is scarce: the weekend at Area A.2 is the busiest setting, while the the weekday at Area B.2 has the least traffic.} 

The findings are notable in comparison to each location's time-to-drive. See Table~\ref{tab:underestimates}. Even for the \text{best} probability-aware policies across all scenarios studied, the average time-to-arrive is 67\% to 123\% higher than the time-to-drive directly to the destinations. This discrepancy is up to 462\% and 173\% for the patient and impatient baselines, respectively. This shows how travel time projections that do not account for parking can substantially underestimate the true value.

This difference may be relevant for users deciding which mode to take. In the case of the route from Kerry Park to Pike Place Market (A.2), the public transit time-to-arrive estimates are substantially shorter than the mean time-to-arrive values of the baseline driving strategies and similar to the values from the best probability-aware driving strategies; see Table~\ref{tab:transit}. In contrast, the time-to-drive projection is 50\% \textit{shorter} than the average public transit time-to-arrive. This could mean that users may mistakenly disregard public transit because they believe it requires double the time of driving a personal vehicle, when the opposite may be true (or the options may at least be similar). Note that uncongested drive time estimates were used for conservatism -- we expect increased traffic would tend to add more time to the parking strategies than the direct drive time, since they generally comprise longer distances.

\begin{table}[tbp]
\caption{Time-to-Arrive vs. Time-to-Drive: \ch{Abs. \& Pct. Differences (\textdownarrow)}}
\label{tab:underestimates}
\begin{tabular*}{\columnwidth}{@{\extracolsep{\fill}}lllcccc}
\toprule
\cmidrule(lr){4-7}
 & & & \multicolumn{2}{c}{10\% Adoption} & \multicolumn{2}{c}{50\% Adoption} \\
\cmidrule(lr){4-5} \cmidrule(lr){6-7}
Dest. & Day & Policy & Minutes & \% & Minutes & \% \\
\midrule
\multirow{6}{*}{\makecell{Area\\A.2\\(Pike\\Place)}} & \multirow{3}{*}{\makecell{\\Week-\\day}} & BL-Pat. & \cellcolor{reddark}34.6 & \cellcolor{reddark}346\% & \cellcolor{reddark}35.2 & \cellcolor{reddark}352\% \\
& & BL-Imp. & \cellcolor{redmed}11.9 & \cellcolor{redmed}119\% & \cellcolor{redmed}10.8 & \cellcolor{redmed}108\% \\
& & Best PA & \cellcolor{redlight}9.0 & \cellcolor{redlight}90\% & \cellcolor{redlight}9.1 & \cellcolor{redlight}91\% \\
\cmidrule{2-7}
& \multirow{3}{*}{\makecell{\\Week-\\end}} & BL-Pat. & \cellcolor{reddark}45.6 & \cellcolor{reddark}456\% & \cellcolor{reddark}46.2 & \cellcolor{reddark}462\% \\
& & BL-Imp. & \cellcolor{redmed}17.3 & \cellcolor{redmed}173\% & \cellcolor{redmed}15.3 & \cellcolor{redmed}153\% \\
& & Best PA & \cellcolor{redmed}11.9 & \cellcolor{redmed}119\% & \cellcolor{redlight}9.1 & \cellcolor{redlight}91\% \\
\midrule
\multirow{6}{*}{\makecell{Area\\B.2\\(Bot.\\Gdn.)}} & \multirow{3}{*}{\makecell{\\Week-\\day}} & BL-Pat. & \cellcolor{redlight}4.3 & \cellcolor{redlight}72\% & \cellcolor{redlight}4.2 & \cellcolor{redlight}70\% \\
& & BL-Imp. & \cellcolor{redlight}4.8 & \cellcolor{redlight}80\% & \cellcolor{redlight}4.5 & \cellcolor{redlight}75\% \\
& & Best PA & \cellcolor{redlight}4.0 & \cellcolor{redlight}67\% & \cellcolor{redlight}4.1 & \cellcolor{redlight}68\% \\
\cmidrule{2-7}
& \multirow{3}{*}{\makecell{\\Week-\\end}} & BL-Pat. & \cellcolor{redmed}8.6 & \cellcolor{redmed}143\% & \cellcolor{redmed}8.0 & \cellcolor{redmed}133\% \\
& & BL-Imp. & \cellcolor{redmed}7.8 & \cellcolor{redmed}130\% & \cellcolor{redmed}8.0 & \cellcolor{redmed}133\% \\
& & Best PA & \cellcolor{redmed}7.4 & \cellcolor{redmed}123\% & \cellcolor{redlight}5.6 & \cellcolor{redlight}93\% \\
\bottomrule
\vspace{0.1mm}
\end{tabular*}
Values represent the absolute gap and percentage increase of time-to-arrive over time-to-drive. Each policy's time-to-arrive is the mean recorded in Table~\ref{tab:policy_performance}. `Best PA' refers to the probability-aware policy with the lowest mean time for a given scenario. Each destination's time-to-drive is as described in Section~\ref{subsec:strat_comparison}.1. Lower is better. \ch{Darker shading indicates more severe time-to-drive underestimates relative to time-to-arrive.}
\end{table}

\begin{table}[t]
\caption{Driving Time-to-Arrive vs. Public Transit Time-to-Arrive for Area A.2 (Pike Place Market): \ch{Abs. \& Pct. Differences}}
\label{tab:transit}
\begin{tabular*}{\columnwidth}{@{\extracolsep{\fill}}llcccc}
\toprule
\cmidrule(lr){3-6}
& & \multicolumn{2}{c}{10\% Adoption} & \multicolumn{2}{c}{50\% Adoption} \\
\cmidrule(lr){3-4} \cmidrule(lr){5-6}
Day & Policy & Minutes & \% & Minutes & \% \\
\midrule
\multirow{3}{*}{Weekday} & BL-Patient & \cellcolor{purpledark}24.6 & \cellcolor{purpledark}123\% & \cellcolor{purpledark}25.2 & \cellcolor{purpledark}126\% \\
 & BL-Impat. & \cellcolor{purplemed}1.9 & \cellcolor{purplemed}10\% & \cellcolor{purplelight}0.8 & \cellcolor{purplelight}4\% \\
 & Best PA & \cellcolor{purplelight}-1.0 & \cellcolor{purplelight}-5\% & \cellcolor{purplelight}-0.9 & \cellcolor{purplelight}-4\% \\
\cmidrule{1-6}
\multirow{3}{*}{Weekend} & BL-Patient & \cellcolor{purpledark}35.6 & \cellcolor{purpledark}178\% & \cellcolor{purpledark}36.2 & \cellcolor{purpledark}181\% \\
 & BL-Impat. & \cellcolor{purplemed}7.3 & \cellcolor{purplemed}37\% & \cellcolor{purplemed}5.3 & \cellcolor{purplemed}27\% \\
 & Best PA & \cellcolor{purplemed}1.9 & \cellcolor{purplemed}10\% & \cellcolor{purplelight}-0.9 & \cellcolor{purplelight}-4\% \\
\bottomrule
\vspace{0.1mm}
\end{tabular*}
Values represent the difference in time-to-arrive for driving relative to public transit. Each driving policy's time-to-arrive is defined as in Table~\ref{tab:underestimates}. The public transit time-to-arrive estimate used is 20 minutes, the midpoint of the range described in the Section~\ref{subsec:strat_comparison}.1 site description for Area A.2. For reference, this site's time-to-\textit{drive} estimate is 10 minutes \textit{shorter} (-50\%) than the estimated public transit time-to-arrive. \ch{Darker shading indicates larger deviations. Purple is used since larger values are not necessarily worse -- more accurate total travel time estimates enable fairer cross-mode comparison.}
\end{table}

\subsubsection{\chold{Error impact}} 
\chold{Interestingly, performance is comparable across the two adoption rates. This suggests limited operational impact from errors incurred via stochastic observation. The probability-aware policies' performance relative to their oracles also supports this conclusion, although those at the 50\% adoption rate perform slightly better relative to their oracles than those at the 10\% rate. One explanation for the system's resilience to errors may be the following trade-off: probability estimates (and by extension, errors) may matter more in competitive settings, but competitive parking environments also have higher traffic, which produces more frequent observations of a lot's parking availability. The results across the 1-, 2-, and 3-step lookahead probability-aware policies suggest that deeper lookaheads -- which incur computational complexity exponential in the depth of the search -- may not be necessary.}

\section{Future work}
\label{sec:future_work}

Future work could find analytical expressions for alternative cases or explore uncertainty in drive and walk time estimates. \chold{Large-scale} experiments assessing the advantages and disadvantages of the strategies outlined above would be welcome, as would identification of parking clusters $\mathcal{C}$ \ch{and investigation of additional probability-aware strategies, perhaps with reinforcement learning}. Related work could be done to estimate the reduction in carbon emissions and time saved via mode shift if time-to-drive estimates are corrected to time-to-arrive estimates. These could also integrate driver preference models that consider factors beyond travel time.

\bibliographystyle{IEEEtran}
\bibliography{References}

@article{arnott2017cruising,
  title={Cruising for parking around a circle},
  author={Arnott, Richard and Williams, Parker},
  journal={Transportation research part B: methodological},
  volume={104},
  pages={357--375},
  year={2017},
  publisher={Elsevier}
}

@inproceedings{arora2019hard,
  title={Hard to park? {E}stimating parking difficulty at scale},
  author={Arora, Neha and Cook, James and Kumar, Ravi and Kuznetsov, Ivan and Li, Yechen and Liang, Huai-Jen and Miller, Andrew and Tomkins, Andrew and Tsogsuren, Iveel and Wang, Yi},
  booktitle={Proceedings of the 25th ACM SIGKDD International Conference on Knowledge Discovery \& Data Mining},
  pages={2296--2304},
  year={2019}
}

@article{bertsimas2019travel,
  title={Travel time estimation in the age of big data},
  author={Bertsimas, Dimitris and Delarue, Arthur and Jaillet, Patrick and Martin, S{\'e}bastien},
  journal={Operations Research},
  volume={67},
  number={2},
  pages={498--515},
  year={2019},
  publisher={INFORMS}
}

@article{bock2019smart,
  title={Smart parking: Using a crowd of taxis to sense on-street parking space availability},
  author={Bock, Fabian and Di Martino, Sergio and Origlia, Antonio},
  journal={IEEE Transactions on Intelligent Transportation Systems},
  volume={21},
  number={2},
  pages={496--508},
  year={2019},
  publisher={IEEE}
}

@inproceedings{djuric2016parkassistant,
  title={Parkassistant: An algorithm for guiding a car to a parking spot},
  author={Djuric, Nemanja and Grbovic, Mihajlo and Vucetic, Slobodan},
  booktitle={Transportation Research Board 95th Annual Meeting},
  volume={16},
  pages={5433},
  year={2016}
}

@article{fosgerau2012valuing,
  title={Valuing travel time variability: Characteristics of the travel time distribution on an urban road},
  author={Fosgerau, Mogens and Fukuda, Daisuke},
  journal={Transportation Research Part C: Emerging Technologies},
  volume={24},
  pages={83--101},
  year={2012},
  publisher={Elsevier}
}

@inproceedings{hedderich2018optimization,
  title={Optimization of a Park Spot Route based on the {A}* Algorithm},
  author={Hedderich, Mareike and Fastenrath, Ulrich and Bogenberger, Klaus},
  booktitle={2018 21st International Conference on Intelligent Transportation Systems (ITSC)},
  pages={3493--3498},
  year={2018},
  organization={IEEE}
}

@article{kotb2016iparker,
  title={i{P}arker—{A} new smart car-parking system based on dynamic resource allocation and pricing},
  author={Kotb, Amir O and Shen, Yao-Chun and Zhu, Xu and Huang, Yi},
  journal={IEEE transactions on intelligent transportation systems},
  volume={17},
  number={9},
  pages={2637--2647},
  year={2016},
  publisher={IEEE}
}

@article{krapivsky2019simple,
  title={Simple parking strategies},
  author={Krapivsky, PL and Redner, S},
  journal={Journal of Statistical Mechanics: Theory and Experiment},
  volume={2019},
  number={9},
  pages={093404},
  year={2019},
  publisher={IOP Publishing}
}

@article{ogulenko2022nature,
  title={The nature of the on-street parking search},
  author={Ogulenko, Aleksey and Benenson, Itzhak and Fulman, Nir},
  journal={Transportation Research Part B: Methodological},
  volume={166},
  pages={48--68},
  year={2022},
  publisher={Elsevier}
}

@article{peer2012prediction,
  title={Prediction of travel time variability for cost-benefit analysis},
  author={Peer, Stefanie and Koopmans, Carl C and Verhoef, Erik T},
  journal={Transportation Research Part A: Policy and Practice},
  volume={46},
  number={1},
  pages={79--90},
  year={2012},
  publisher={Elsevier}
}

@article{rizvi2018aspire,
  title={Aspire: An agent-oriented smart parking recommendation system for smart cities},
  author={Rizvi, Syed R and Zehra, Susan and Olariu, Stephan},
  journal={IEEE Intelligent Transportation Systems Magazine},
  volume={11},
  number={4},
  pages={48--61},
  year={2018},
  publisher={IEEE}
}

@article{sakaguchi1982optimal,
  title={On the optimal parking problem in which spaces appear randomly},
  author={Sakaguchi, Minoru and Tamaki, Mitsushi},
  year={1982},
  publisher={Research Association of Statistical Sciences}
}

@misc{seattle_paid_parking_occupancy,
  title = {Paid Parking Occupancy - {L}ast 30 Days},
  author = {Seattle Department of Transportation (SDOT)}, 
  year = {2025},
  url = {https://data.seattle.gov/Transportation/Paid-Parking-Occupancy-Last-30-Days-/rke9-rsvs/about_data},
  urldate = {2025-02-03},
}

@misc{seattle_paid_parking_transactions,
  title = {Paid Parking Transaction Data},
  author = {Seattle Department of Transportation (SDOT)},
  year = {2025},
  url = {https://data.seattle.gov/Transportation/Paid-Parking-Transaction-Data/gg89-k5p6/about_data},
  urldate = {2025-02-03},
}

@article{shi2018parkcrowd,
  title={ParkCrowd: Reliable crowdsensing for aggregation and dissemination of parking space information},
  author={Shi, Fengrui and Wu, Di and Arkhipov, Dmitri I and Liu, Qiang and Regan, Amelia C and McCann, Julie A},
  journal={IEEE Transactions on Intelligent Transportation Systems},
  volume={20},
  number={11},
  pages={4032--4044},
  year={2018},
  publisher={IEEE}
}

@article{shoup2006cruising,
  title={Cruising for parking},
  author={Shoup, Donald C},
  journal={Transport policy},
  volume={13},
  number={6},
  pages={479--486},
  year={2006},
  publisher={Elsevier}
}

@article{tamaki1982optimal,
  title={An Optimal Parking Problem},
  author={Tamaki, Mitsushi},
  journal={Journal of Applied Probability},
  pages={803--814},
  year={1982},
  publisher={JSTOR}
}

@article{tamaki1988optimal,
  title={Optimal stopping in the parking problem with {U}-turn},
  author={Tamaki, Mitsushi},
  journal={Journal of applied probability},
  volume={25},
  number={2},
  pages={363--374},
  year={1988},
  publisher={Cambridge University Press}
}

@article{teodorovic2006intelligent,
  title={Intelligent parking systems},
  author={Teodorovi{\'c}, Du{\v{s}}an and Lu{\v{c}}i{\'c}, Panta},
  journal={European Journal of Operational Research},
  volume={175},
  number={3},
  pages={1666--1681},
  year={2006},
  publisher={Elsevier}
}

@article{uchida2014estimating,
  title={Estimating the value of travel time and of travel time reliability in road networks},
  author={Uchida, Kenetsu},
  journal={Transportation Research Part B: Methodological},
  volume={66},
  pages={129--147},
  year={2014},
  publisher={Elsevier}
}

@inproceedings{wang2018will,
  title={When will you arrive? {E}stimating travel time based on deep neural networks},
  author={Wang, Dong and Zhang, Junbo and Cao, Wei and Li, Jian and Zheng, Yu},
  booktitle={Proceedings of the AAAI conference on artificial intelligence},
  volume={32},
  number={1},
  year={2018}
}

@article{wu2014agile,
  title={Agile urban parking recommendation service for intelligent vehicular guiding system},
  author={Wu, Eric Hsiao-Kuang and Sahoo, Jagruti and Liu, Chi-Yun and Jin, Ming-Hui and Lin, Shu-Hui},
  journal={IEEE Intelligent Transportation Systems Magazine},
  volume={6},
  number={1},
  pages={35--49},
  year={2014},
  publisher={IEEE}
}

@article{xiao2023parking,
  title={Parking prediction in smart cities: A survey},
  author={Xiao, Xiao and Peng, Ziyan and Lin, Yunqing and Jin, Zhiling and Shao, Wei and Chen, Rui and Cheng, Nan and Mao, Guoqiang},
  journal={IEEE Transactions on Intelligent Transportation Systems},
  volume={24},
  number={10},
  pages={10302--10326},
  year={2023},
  publisher={IEEE}
}

@article{xiao2018likely,
  title={How likely am {I} to find parking? -- {A} practical model-based framework for predicting parking availability},
  author={Xiao, Jun and Lou, Yingyan and Frisby, Joshua},
  journal={Transportation Research Part B: Methodological},
  volume={112},
  pages={19--39},
  year={2018},
  publisher={Elsevier}
}

@book{newell1982,
  title={Applications of Queueing Theory},
  author={G. F. Newell},
  publisher={Springer},
  year={1982}
}

@article{Jorge2012,
author = {Jorge A. Laval and Zhengbing He and Felipe Castrillon},
title ={Stochastic Extension of {N}ewell's Three-Detector Method},
journal = {Transportation Research Record},
volume = {2315},
number = {1},
pages = {73-80},
year = {2012}
}

@book{kutoyants2023,
  title={Introduction to the Statistics of Poisson Processes and Applications},
  author={Yury A. Kutoyants},
  publisher={Springer Nature},
  year={2023}
}

@misc{Yafeng2024,
author = {Yafeng Yin},
title = {Private communication during Conference in Emerging Technologies in Transportation Systems ({TRC}-30)},
year = {2024}
}

\vspace{-10mm}

\begin{IEEEbiography}[{\vspace{-6mm}\includegraphics[width=1in,height=1.25in,clip,keepaspectratio]{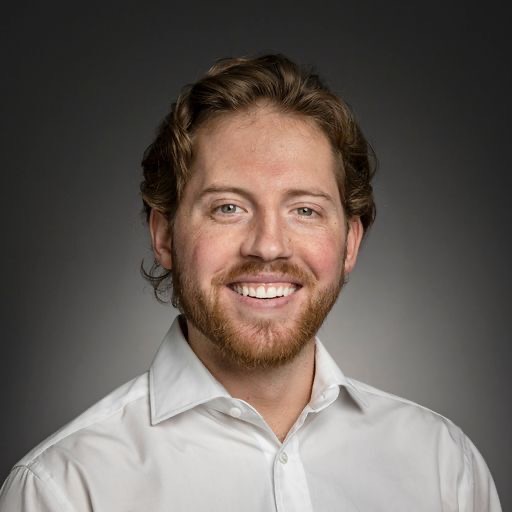}}] {Cameron Hickert} earned a B.S. in physics from the University of Denver and an M.E. in computational science and engineering from Harvard University. He is currently a Ph.D. student in Social and Engineering Systems at MIT. His research interests include safe autonomy, reinforcement learning, and distributed cyber-physical systems.
\end{IEEEbiography}

\vspace{-10mm}

\begin{IEEEbiography}[{\vspace{-3mm}\includegraphics[width=1in,height=1.25in,clip,keepaspectratio]{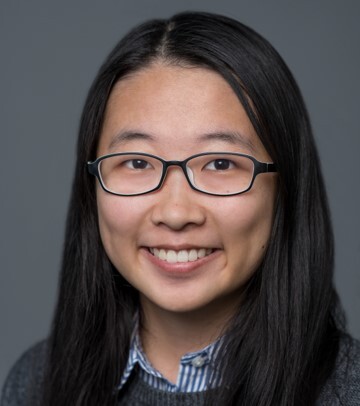}}]{Sirui Li} received her B.S. degree with majors in computer science and mathematics from Washington University in St. Louis, MO, USA, in 2019. She is currently working toward her Ph.D. degree in Social and Engineering Systems at MIT, Cambridge, MA, USA. Her research interests include areas of machine learning for combinatorial optimization and control analysis for transportation systems.
\end{IEEEbiography}

\vspace{-10mm}

\begin{IEEEbiography}[{\includegraphics[width=1in,height=1.25in,clip,keepaspectratio]{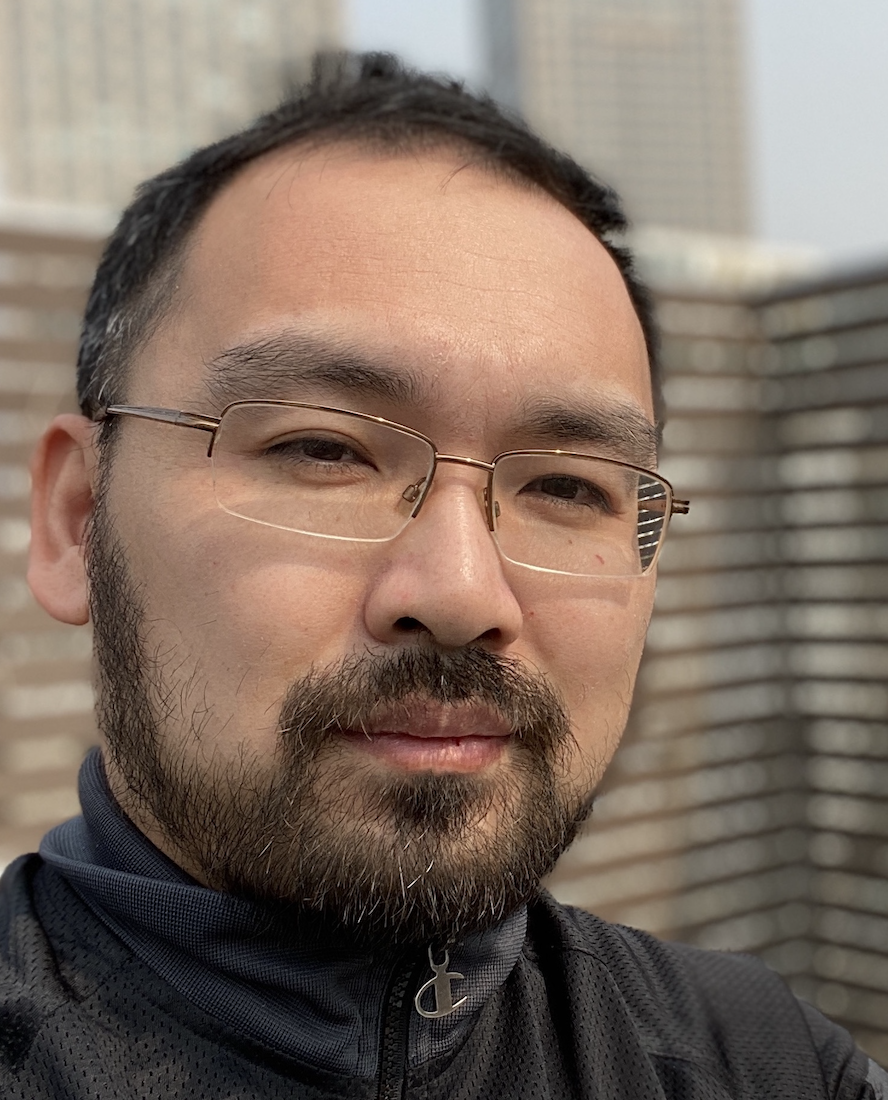}}] {Zhengbing He} (M'17-SM'20) received the Bachelor of Arts degree in English language and literature from Dalian University of Foreign Languages, China, in 2006, and the Ph.D. degree in systems engineering from Tianjin University, China, in 2011. He was a Post-Doctoral Researcher and an Assistant Professor with Beijing Jiaotong University, China. From 2018 to 2022, he was a Full Professor with Beijing University of Technology, China. Presently, he is a Research Scientist at MIT. 

His research stands at the intersection of transportation, systems engineering, and artificial intelligence. He has published more than 180 papers, with total citations exceeding 7,000. He is the Editor-in-Chief of Journal of Transportation Engineering and Information (Chinese). Meanwhile, he serves as a Senior Editor for IEEE TRANSACTIONS ON INTELLIGENT TRANSPORTATION SYSTEMS, an Associate Editor for IEEE TRANSACTIONS ON INTELLIGENT VEHICLES, and an Editorial Advisory Board Member for Transportation Research Part C. His webpage is https://www.GoTrafficGo.com.
\end{IEEEbiography}

\vspace{-10mm}

\begin{IEEEbiography}[{\vspace{-6mm}\includegraphics[width=1in,height=1.25in,clip,keepaspectratio]{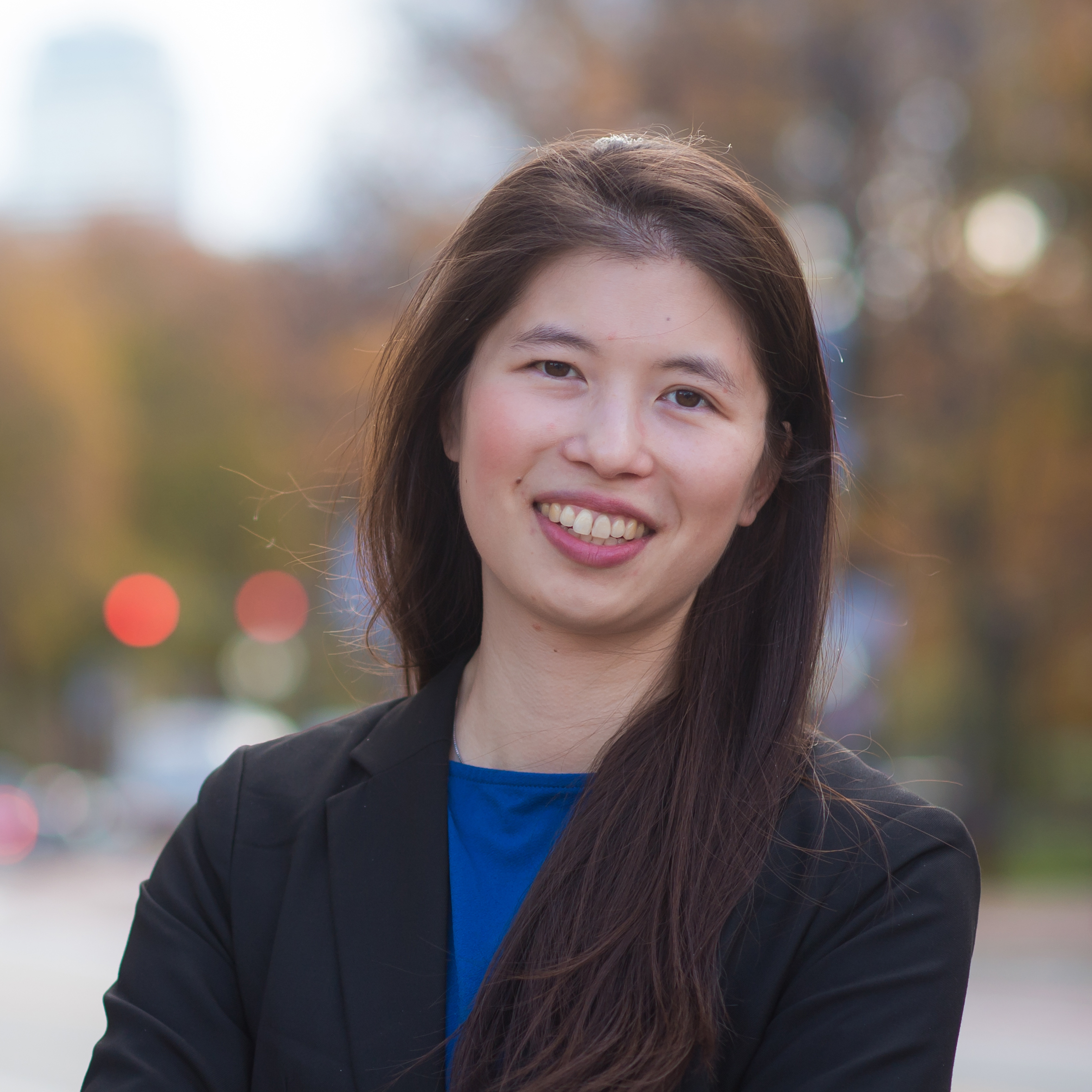}}]{Cathy Wu} (Member, IEEE) received B.S. and M.Eng. degrees from MIT and a Ph.D. degree from UC Berkeley, all in EECS. She was a post-doctoral researcher at Microsoft Research. She is currently the Class of 1954 Career Development Associate Professor at MIT in LIDS, CEE, and IDSS. She studies machine learning for optimization, with a focus on urban mobility. She is interested in enabling faster, evidence-driven decisions for sociotechnical systems. Cathy is the recipient of the NSF CAREER (2023), the Ole Madsen Mentoring Award (2025), the IEEE ITS Best Dissertation Award (2019), and the CUTC Milton Pikarsky Memorial Award (2018). She serves on the Board of Governors for the IEEE ITSS, is an Associate Editor or Area Chair for ICML, NeurIPS, ICRA, and Transportation Research Part C, and served as Program Co-chair for RLC 2025. She is also the inaugural Chair and Co-founder of the REproducible Research In Transportation Engineering (RERITE) Working Group.
\end{IEEEbiography}

\vfill

\end{document}